\documentclass[10pt,preprint]{aastex} 

\usepackage{lscape}

\newcommand{\ovi}{O$\;${\small\rm VI}\relax}
\newcommand{\civ}{C$\;${\small\rm IV}\relax}

\newcommand{\cii}{C$\;${\small\rm II}\relax}

\newcommand{\ari}{Ar$\;${\small\rm I}\relax}
\newcommand{\feii}{Fe$\;${\small\rm II}\relax}
\newcommand{\oi}{O$\;${\small\rm I}\relax}

\newcommand{\ntwo}{N$\;${\small\rm II}\relax}
\newcommand{\cthree}{C$\;${\small\rm III}\relax}
\newcommand{\fethree}{Fe$\;${\small\rm III}\relax}
\newcommand{\sthree}{S$\;${\small\rm III}\relax}

\newcommand{\htwo}{H$_2$}

\newcommand{\halpha}{H$\alpha$}

\newcommand{\HI}{H$\;${\small\rm I}\relax}
\newcommand{\HII}{H$\;${\small\rm II}\relax}
\newcommand{\lya}{Lyman-$\alpha$}

\newcommand{\msun}{M$_\odot$}
\newcommand{\kms}{km~s$^{-1}$\relax}

\newcommand{\mvlsr}{v_{\rm LSR}\relax}
\newcommand{\percc}{cm$^{-3}$\relax}
\newcommand{\column}{cm$^{-2}$}
\newcommand{\bval}{$b$-value}
\newcommand{\nav}{$N_a(v)$}

\newcommand{\fuse}{{\em FUSE}}

\newcommand{\iue}{{\em IUE}}

\newcommand{\vz}{vZ~1128}

\begin{document}

\slugcomment{\bf Accepted by the {\em ApJ}}

\title{Ionized Gas in the First 10 kpc of the Interstellar Galactic Halo}

\author{J. Christopher Howk\altaffilmark{1,2}, 
  Kenneth R. Sembach\altaffilmark{3},
  \& Blair D. Savage\altaffilmark{4}}

\altaffiltext{1}{Department of Physics and Astronomy,
  The Johns Hopkins University, Baltimore, MD, 21218}

\altaffiltext{2}{Current Address: Center for Astrophysics and Space
  Sciences, University of California at San Diego, C-0424, La Jolla,
  CA, 92093; howk@trafalgar.ucsd.edu}

\altaffiltext{3}{Space Telescope Science Institute, 
  Baltimore, MD, 21218; sembach@stsci.edu}

\altaffiltext{4}{Astronomy Department, University of Wisconsin-Madison,
  Madison, WI,  53711; savage@astro.wisc.edu}


\begin{abstract}
  
  We present {\em Far Ultraviolet Spectroscopic Explorer} (\fuse)
  observations of the post-asymptotic giant branch star von Zeipel
  1128 ($l=42\fdg5, \, b=+78\fdg7; \ d=10.2 \ {\rm kpc}, z=10.0 \ {\rm
    kpc}; \ v_{\rm LSR} = -140\pm8$ \kms), located in the globular
  cluster Messier 3.  The \fuse\ observations cover the wavelength
  range 905 -- 1187 \AA\ at $\sim20$ \kms\ (FWHM) resolution.  These
  data exhibit many photospheric and interstellar absorption lines,
  including absorption from ions associated with the warm neutral,
  warm ionized, and highly-ionized phases of the interstellar medium
  along this sight line.  We derive interstellar column densities of
  \HI, \ion{P}{2}, \ari, \feii, \fethree, \sthree, and \ovi, with
  lower limits for \ion{C}{2}, \ion{C}{3}, \ion{N}{1}, \oi, and
  \ion{Si}{2}.  Though the individual velocity components within the
  absorption profiles are unresolved by \fuse, a comparison of the
  velocity distribution of depleted or ionized species with the
  neutral species suggests the thick disk material along this sight
  line is infalling onto the Galactic plane, while material near the
  plane is seen closer to rest velocities.  Ionized hydrogen
  represents $\ga12\%$, most likely $\sim45\%$, of the total hydrogen
  column along this sight line, most of it associated with the warm
  ionized phase.  The warm ionized and neutral media toward von Zeipel
  1128 have very similar gas-phase abundances and kinematics: the
  neutral and ionized gases in this region of the thick disk are
  closely related.  Strong \ovi\ absorption is seen with the same
  central velocity as the warm ionized gas, though the \ovi\ velocity
  dispersion is much higher ($\sigma \equiv \sqrt{2}b = 32$ \kms).
  Virtually all of the \ovi\ is found at velocities where
  lower-ionization gas is seen, suggesting the \ovi\ and WNM/WIM
  probes are tracing different portions of the same structures (e.g.,
  the \ovi\ may reside in interfaces surrounding the WNM/WIM clouds).
  We see no evidence for interstellar absorption associated with the
  globular cluster Messier 3 itself nor with the circumstellar
  environment of von Zeipel 1128.  Neither high velocity cloud
  absorption (with $|\mvlsr| \ga 125$ \kms) nor high
  velocity-dispersion gas (with $\sigma \sim60$ \kms) is seen toward
  von Zeipel 1128.

\end{abstract}



\section{Introduction}
\label{sec:intro}

There is now substantial evidence that the interstellar medium (ISM)
of the Milky Way is best described as a ``multiphase'' medium.  In
standard equilibrium models of the ISM (e.g., McKee \& Ostriker 1977,
Wolfire et al. 1995), the term ``phase'' refers to gaseous regions of
distinct temperatures, densities, and ionization states.  Most models
assume these regions are physically independent and exist in
(sometimes rough) pressure equilibrium.  Extraplanar material in
spiral galaxies seems to exhibit all of the principal phases found in
the disk of the Milky Way: the cold neutral medium [CNM; $T \sim 10^2$
K; $x(\mbox{\HII}) \equiv n(\mbox{\HII})/n({\rm H_{tot}}) \sim
10^{-4}$; Howk \& Savage 1999a, 2000; Richter 2001]; the warm neutral
medium [WNM; $T\sim10^4$ K; $x(\mbox{\HII}) \sim 0.1$; Dickey \&
Lockman 1990]; the warm ionized medium [WIM; $x(\mbox{\HII}) \ga 0.8$;
Reynolds 1993]; and the hot ISM [$T\sim 10^6$ K; $x(\mbox{\HII}) \sim
1.0$; Snowden et al. 1998].

While the evidence strongly suggests their presence in the extraplanar
regions of the Milky Way, the properties and physics of these phases
and their relationships are as yet poorly characterized for
extraplanar gas.  As such, there are many unanswered questions
regarding the nature of gas in the thick disk\footnote{In this paper
  we will often use the term ``thick disk'' to refer to the
  extraplanar distribution of interstellar matter in the Milky Way.
  This is analogous to the term ``halo'' used by many authors.  We
  make this choice of terminology both because the ISM within the
  first few kiloparsecs of the midplane tends to rotate as a disk and
  because we wish to distinguish this lower-lying material from the
  much more extended corona, whose presence is inferred by
  observations of \ovi\ associated with very distant gas clouds (e.g.,
  high-velocity clouds; see Sembach et al.  2000, 2003). } of the
Milky Way and other spiral galaxies.  What fraction of the
interstellar thick disk is associated with each of these phases?  What
are the physical conditions within each of these phases?  How does gas
far from the midplane differ from that in the interstellar thin disk?
What implications can be drawn from a comparison of the conditions of
gas associated with the thin and thick disks?  Are the identified
phases simply artifacts of the manner in which the gas of the Galactic
thick disk is studied?  Similarly, are the various tracers of thick
disk matter probing different portions of the same structures?

While there is still relatively little information available to answer
such questions, the relationship between the warm neutral and warm
ionized gas of the thick disk is the best studied.  Spitzer \&
Fitzpatrick's (1993) study of the sight line to the distant star HD
93521 suggests a strong relationship between the warm neutral and
ionized phases of the Galactic thick disk.  They used high-resolution
absorption line spectroscopy to show that excited-state \cii\ (i.e.,
\cii$^\ast$), produced by collisions of ground-state \cii\ with warm
electrons, exhibits the same velocity structure as other
low-ionization species thought to trace the WNM (e.g., \feii,
\ion{Si}{2}, and \ion{S}{2}).  Spitzer \& Fitzpatrick argued that the
electrons and the neutral hydrogen were well-mixed in a
partially-ionized gas.  Howk \& Savage (1999b) found a similar
kinematic alignment for the sight line toward $\rho$ Leo (HD~91316),
which lies near HD~93521 on the sky.  Along this sight line absorption
from low-ionization lines such as \ion{Zn}{2} were shown to have a
velocity structure very similar to that seen in the moderately-ionized
species \sthree\ and \ion{Al}{3}.  The latter ions have ionization
potentials significantly higher than that of \HI, making them unlikely
to occur in regions dominated by neutral hydrogen.

Reynolds et al. (1995) approached this issue by comparing the
distribution of \halpha\ and \HI\ 21-cm emission over a
$10^\circ\times12^\circ$ region of the sky centered on
$(l,b)=(144^\circ,-21^\circ)$.  They noted the existence of ``\halpha
-emitting \HI\ clouds,'' structures seen in both the \halpha\ and \HI\ 
maps that share morphological and kinematical properties when viewed
at the low spatial resolution of their typical maps.  They showed the
ratio of \halpha\ intensity to \HI\ column density increases with
height above the plane in this direction.  While Reynolds et al. found
several prominent examples of structures that seem to bear both
neutral and ionized hydrogen, when they examined one such filament at
higher spatial resolution, they found that the neutral and ionized
gases tended to be displaced from one another on the sky.  While the
\halpha\ and \HI\ were associated with the same filament, the ionized
and neutral gases traced different portions of this structure.

If these examples are representative of the high-latitude sky, they
suggest much of the gas associated with the neutral and ionized phases
of the Galactic thick disk are distributed in a similar manner with
similar physical conditions.  If the Spitzer \& Fitzpatrick
proposition that neutral and ionized hydrogen are intermixed in the
same structures is true, then the ionization of hydrogen (and the
\sthree\ and \ion{Al}{3} toward $\rho$ Leo) within these ``clouds'' 
most likely occurs by means other than photoionization by OB stars
(e.g., shocks, X-ray photoionization, or cosmic-ray ionization),
although it is generally thought that only ionizing radiation from OB
stars has enough power to provide for the observed ionization of the
WIM.  Furthermore, the ionization of hydrogen through mechanisms such
as shocks and X-ray photoionization may also produce by-products such
as the highly-ionized gas traced by \ovi\ and \civ.  The work of
Reynolds et al. (1995) suggests a close relationship between the warm
neutral and ionized gases, though one in which these phases are indeed
distinct.

These studies are all difficult to generalize to the whole
high-latitude sky.  Too few sight lines have been studied in the
detailed manner that the three discussed above have been to draw
conclusions about the general relationship between the phases of the
Galactic thick disk.  Furthermore, these studies provide little
insight into the role of hotter, more highly-ionized material.

This paper presents {\em Far Ultraviolet Spectroscopic Explorer}
(\fuse) observations of the post-asymptotic giant branch (post-AGB)
star von Zeipel 1128 (hereafter \vz; von Zeipel 1908).  This star lies
in the globular cluster Messier 3 (NGC 5272), located 10.2 kpc from
the Sun at a height above the Galactic midplane of 10.0 kpc (Harris
1996).  Thus, the sight line to \vz\ probes nearly all of the thick
disk and much of the corona of the Galaxy, including the most distant
reaches typically only accessible via observations of extragalactic
targets (e.g., Savage et al. 2000; Savage et al. 2003).

We use the present \fuse\ observations to study the content and
abundances of the ionized gas, as well as its kinematics, along this
extended Galactic sight line.  We will emphasize the ionization
characteristics of the sight line and the connection between the
various ``phases'' of the ISM in this direction.

In \S \ref{sec:sightline} we discuss the properties of the star and
sight line.  We summarize the \fuse\ observations and our reduction of
the data in \S \ref{sec:observations}, while we summarize our analysis
methodology in \S \ref{sec:analysis}.  We present the properties of
the highly-ionized gas along the \vz\ sight line in \S \ref{sec:osix}
and of the warm ionized and neutral media along this path in \S
\ref{sec:warm}.  In \S \ref{sec:highvelocity} we detail the lack of
high-velocity absorption along this sight line.  In \S
\ref{sec:discussion} we discuss these results, and we summarize our
principle conclusions in \S \ref{sec:summary}.

\section{Stellar and Interstellar Sight Line Properties}
\label{sec:sightline}

Table \ref{tab:star} summarizes the properties of the star \vz\ and
the interstellar path to this object.  First cataloged by von Zeipel
(1908), \vz\ (also known as NGC 5272 ZNG 1, K 728, II-57) is
classified as a hot ($T_{\rm eff} \approx 33,000$ K; Dixon et al.
1994) post-AGB star.  Such objects are thought to be the remnants of
AGB stars that have ceased nucleosynthetic burning.  By this stage of
evolution, stars have shed significant amounts of mass (i.e., their
complete outer envelopes) and are on their way to becoming white
dwarfs.  Not all post-AGB stars pass through a planetary nebula phase.
Lower mass objects leave the AGB before the thermal pulsing phase of
AGB evolution.  These low-mass stars evolve very slowly (by the
standards of higher-mass post-AGB stars); by the time they are hot
enough to provide significant ionizing flux, the material that would
have become a planetary nebula has since dispersed (see review by
Moehler 2001).

Relatively few observations of \vz\ have been reported in the
literature.  Studies of the stellar properties include works by de
Boer (1985) using the {\em International Ultraviolet Explorer} (\iue)
and Dixon et al. (1994) using the {\em Hopkins Ultraviolet Telescope}.
The stellar properties arising from these studies place \vz\ on
theoretical tracks of post-AGB evolution (e.g., Sch\"{o}nberner 1983)
with a mass of $\sim0.55$ \msun\ (Dixon et al. 1994), though model
atmosphere fits give slightly smaller values.  The radial velocity
reported for this star in Table \ref{tab:star} is from the present
work.  The quoted error is dominated by the uncertainties in the
absolute velocity calibration of \fuse\ (discussed below).


The interstellar sight line to \vz\ traces a pathlength of $\sim10.2$
kpc almost perpendicular to the Galactic plane (Djorgovski 1993).
This extended path samples gas in the northern half of the Galactic
thin disk, thick disk, and extended corona.  Given the high latitude
of this sight line, the path through the thin disk of the Milky Way is
quite short, approximately the half-thickness of this layer.  Much of
the absorption observed along this sight line is therefore expected to
be from material significantly above the Galactic plane.  The
interstellar sight line to \vz\ has previously been discussed by de
Boer \& Savage (1984), who note the presence of negative velocity
absorption in a high-resolution (low signal to noise) {\em IUE}
spectrum.  They comment that this material may be associated with the
infall of matter participating in a ``galactic fountain.''

\begin{figure}  
\epsscale{0.7}
\plotone{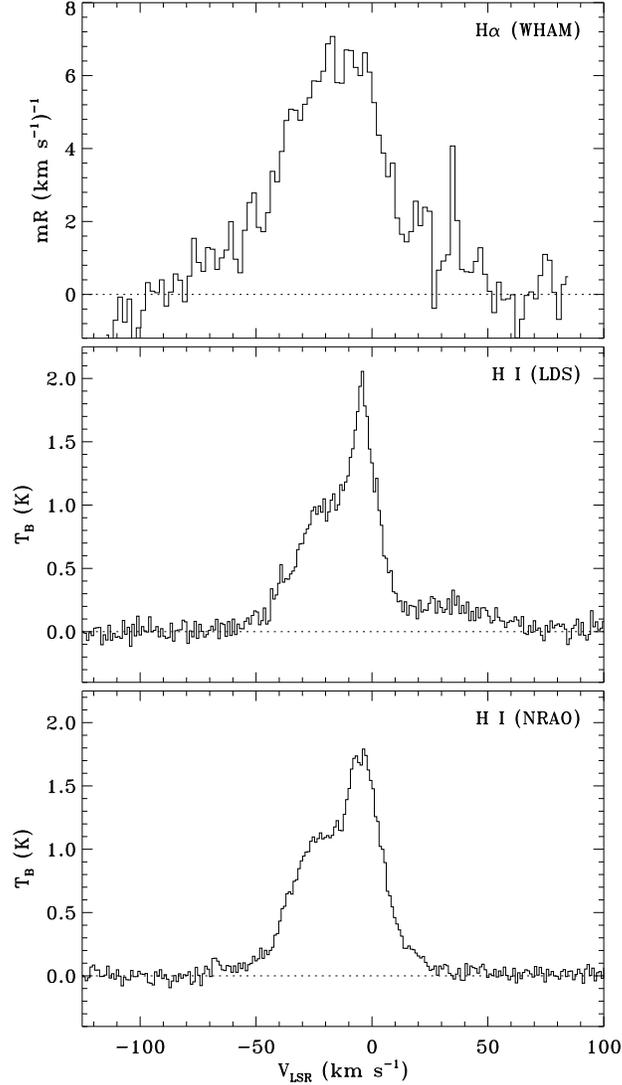}
\caption{Ground-based spectra of ionized and neutral hydrogen emission 
  towards vZ~1128.  The top panel is a WHAM spectrum of \halpha\ 
  emission taken with a $1^\circ$ beam at $\sim12$ \kms\ resolution
  (Haffner et al.  2002).  The spectrum shown here is an average of
  all pointings within $1\fdg5$ of vZ~1128 and therefore probes the
  \halpha\ emission of a region of the sky with radius $\sim2^\circ$.
  The middle spectrum is from the Leiden-Dwingeloo \HI\ 21-cm Survey
  (Hartmann \& Burton 1997).  This survey has a 30\arcmin\ beam and
  1.0 \kms\ resolution.  The bottom spectrum is an NRAO 140-ft \HI\ 
  21-cm spectrum from Danly et al. (1992).  These data sample a beam
  size of 21\arcmin\ at $1.0$ \kms\ resolution.
\label{fig:hydrogenspectra}}
\end{figure}

Figure \ref{fig:hydrogenspectra} shows emission line profiles for
ionized and neutral hydrogen in the general direction of \vz.  The top
panel shows the \halpha\ spectrum towards \vz\ from the Wisconsin
H-alpha Mapper (WHAM; Haffner et al. 2002; Reynolds et al. 1998b) and
represents an average of the eight pointings centered within $1\fdg5$
of \vz.  WHAM has a $1^\circ$ beam and 12 \kms\ velocity resolution.
The next two panels show \HI\ 21-cm spectra from the Leiden-Dwingeloo
Survey (LDS; Hartmann \& Burton 1997) with a 30\arcmin\ beam and from
the NRAO 140-ft telescope (unpublished from the Danly et al. 1992
survey) with a 21\arcmin\ beam.  The LDS spectrum is centered
$\sim12\arcmin$ from \vz\ at $(l,b) = (42\fdg5, \, +78\fdg5)$.  Both
\HI\ spectra have a velocity resolution of $\sim1$ \kms\ and have been
cleaned of stray radiation.  The weak positive velocity wing (centered
at $\mvlsr \approx +30$ \kms) in the LDS spectrum may result from
imperfect baseline or stray radiation correction.

These spectra demonstrate the presence of neutral and ionized material
over the velocity range $-70\la \mvlsr \la +30$ \kms.  The total \HI\
column densities over this range measured by the LDS and NRAO spectra
are $N(\mbox{\HI}) = 1.0\times10^{20}$ and $1.1\times10^{20}$ \column,
respectively.  An analysis of the \lya\ absorption in recent Space
Telescope Imaging Spectrograph echelle-mode data suggest
$N(\mbox{\HI})=(9.3\pm0.7)\times10^{19}$ \column\ (Howk, Sembach, \&
Savage 2003).  The difference in these column densities is likely a
result of \HI\ structure on scales smaller than the beam size of the
radio observations.

The \HI\ emission spectra show a strong peak near $\mvlsr \sim -4$
\kms, which is close to the expected location of gas in the Galactic
thin disk or material in the halo that is corotating with the disk.
The next most prominent component or blend of components is centered
near $\mvlsr \sim -25$ \kms.  These two blends have nearly equal
column densities and account for $99\%$ of the total
\HI\ column in the NRAO spectrum.

There is little evidence for significant amounts of cold material
along this sight line in the \HI\ spectrum.  While Gaussian
decomposition of the \HI\ line profiles is at best non-unique, we have
found that reasonable fits yield $b\approx10$ \kms\ for the strongest
component centered at $\mvlsr \sim -4$ \kms, where $b$ is the Doppler
parameter [$b\equiv (2kT/m)^{1/2}$ for pure thermal broadening;
Spitzer 1978].  This corresponds to a kinetic temperature of $T \la
6,000$ K.  (The LDS spectrum yields a slightly smaller $b$-value:
$b\approx7$ \kms.) The gas at more negative velocities can be fitted
by Gaussians with widths $9 \la b \la 15$ \kms\ implying upper limits
on the temperature between $5,000$ K and $14,000$ K.  While a small
amount of cool gas could be blended within the emission profile, most
of the neutral material along this sight line seems to be associated
with the WNM.

The negative-velocity gas is seen more prominently in the \halpha\ 
spectrum than the \HI\ spectra, which in the simplest interpretation
suggests this infalling material has a greater degree of
ionization.\footnote{It should be noted that while the \HI\ spectrum
  is weighted by the neutral hydrogen density, the \halpha\ spectrum
  is weighted by the square of the electron density.  Furthermore, the
  \halpha\ spectrum shown is an average over a field of radius
  $\sim2^\circ$.} The total intensity of the \halpha\ line averaged
over the pointings centered within $1\fdg5$ is 0.39 Rayleighs, which
corresponds to an emission measure of $\approx0.68 \, (T/6000)^{0.9}$
pc cm$^{-6}$.  The intensity of the nearest one degree field to \vz\ 
is $I(\mbox{\halpha})=0.17\pm0.03$, or $EM \approx [0.30\pm0.06] \,
(T/6000)^{0.9}$ pc cm$^{-6}$.  If the mean density of the ionized
material is assumed to be $n_e \sim 0.08$ (see Reynolds 1991a), then
the \HII\ column density along the sight line to \vz\ (averaged over
the $1^\circ$ beam) is $N(\mbox{\HII}) \sim 1.2\times10^{19}$ \column\ 
for $T=6,000$ K ($\sim1.8\times10^{19}$ \column\ for $T=10,000$ K).
In this case, the column density of ionized gas is $\sim13\%$
($\sim19\%$) that of the neutral gas (see also \S
\ref{sec:ionization}).  The average ratio of the column densities
$N(\mbox{\HII})/N(\mbox{\HI})$ integrated vertically through the
Galaxy is between $\sim 20\% \, - \,30\%$ (G\'{o}mez, Benjamin, \& Cox
2001; Reynolds 1993; Dickey \& Lockman 1990).  We note there is large
degree of variation in the \halpha\ intensities in the WHAM
measurements of this direction, with the beam centered closest to \vz\ 
having the lowest intensity of those pointings centered within
$1\fdg5$.  The ratio of ionized to neutral gas can be as high as
$29\%$ ($44\%$ if $T=10,000$ K) assuming an emission measure averaged
over the pointings centered within $1\fdg5$ (see Figure
\ref{fig:hydrogenspectra}).  These estimates of $N(\mbox{\HII})$
should be considered to have large uncertainties since the assumption
of a single (poorly known) average electron density is unlikely to be
correct.  Identifying a pulsar in M3 would be extremely helpful to
this investigation since it would allow the direct determination of
$N(\mbox{\HII})$ through its dispersion measure.

%
%

The direct comparison of these spectra, which sample large regions of
the sky ($\sim0\fdg35-2\fdg0$), with our absorption line measurements
is complicated by the difference in beam size of the observations.
However, the \HI\ and \halpha\ measurements provide information on the
average properties of the neutral and ionized material in this general
direction.  In particular they show evidence for significant
departures from the expected Galactic rotation, since gas along
high-latitude sight lines participating in this rotation should be
seen nearly at rest relative to the Local Standard of Rest (LSR; see
\S \ref{subsec:kinematics}).  A comparison of the \HI\ and \halpha\ 
profiles also reveals the increased importance of ionized gas in the
negative-velocity material compared with that near the LSR.

Hot gas ($T\ga 10^6$ K) in the Galactic halo can be traced through its
soft X-ray emission (Snowden et al. 1997, 1998).  The {\em ROSAT}
1/4-keV count rate averaged over a 36\arcmin\ region toward \vz\ is
$1178\pm142$ counts s$^{-1}$ arcmin$^{-2}$ (Savage et al. 2003).  For
comparison, the average 1/4-keV count rate for latitudes $b\ge
60^\circ$ derived from the maps of Snowden et al.  (1997) is
$1034\pm255$ counts s$^{-1}$ arcmin$^{-2}$, where the standard
deviation is quoted as the uncertainty.  While the interpretation of
soft X-ray count rates is strongly dependent upon the total column
density of absorbing material along the sight line, it seems the count
rate along the sight line toward \vz\ is comparable to the
high-latitude average.

\section{Observations and Reductions}
\label{sec:observations}

Table \ref{tab:log} gives a log of the \fuse\ observations used for
this work.  The planning and in-orbit performance of \fuse\ are
discussed by Moos et al. (2000) and Sahnow et al. (2000).  \fuse\ 
consists of four co-aligned telescopes and Rowland-circle
spectrographs feeding two microchannel plate (MCP) detectors with
helical double delay line anodes.  Two of the telescope/spectrograph
channels have SiC coatings providing reflectivity over the range $\sim
905 - 1105$ \AA, while the other two have Al:LiF coatings for
sensitivity in the $\sim1000 - 1187$ \AA\ range.  We refer to these as
SiC and LiF channels, respectively.

The light reflected from each mirror passes through a focal plane
assembly containing three entrance apertures; all of the \vz\ data
were taken through the $30\arcsec \times 30\arcsec$ (LWRS) apertures.
The four holographically-ruled spherical gratings project the spectra
onto two detectors (each made up of two segments -- segments A and B),
with a LiF and SiC channel projected in parallel onto each detector.
For the \vz\ observations, the data were taken in time-tag mode, so
that each detected photon is recorded by its position and arrival
time.

The individual exposures for the three sets of observations (see Table
\ref{tab:log}) were merged into a single spectrum by concatenating the
individual photon event lists.  We employed the {\tt CALFUSE} (v2.0.5)
pipeline for transforming the photon event lists to a calibrated
one-dimensional spectrum.  {\tt CALFUSE} was used to screen the photon
lists for valid data with constraints imposed for Earth limb angle
avoidance and passage through the South Atlantic Anomaly.  Corrections
for detector backgrounds, Doppler shifts caused by spacecraft orbital
motions, geometrical distortions, and astigmatism were applied (Sahnow
et al. 2000). The {\tt CALFUSE} processing is described in more detail
by Oegerle et al. (2000).

The processed data have a nominal spectral resolution of $\la20$ \kms\ 
(FWHM), with a $1\sigma$ relative wavelength dispersion solution
accuracy of $\la \pm6$ \kms.  The zero point of the wavelength scale
for individual \fuse\ observations is poorly determined.  For this
work we compare the \HI\ 21-cm line observations of Danly et al.
(1992) with the absorption profiles of \ari\ $\lambda1048.220$ and
$\lambda1066.660$ to correct the \fuse\ LiF1A data to the local
standard of rest.\footnote{We adopt the standard definition of the
  LSR, assuming a solar motion of $+20$ \kms\ in the direction
  $(\alpha,\delta)_{1900} = (18^h,+30^\circ)$
  [$(l,b)\approx(56^\circ,+23^\circ)$].  This gives $\mvlsr =
  v_{helio}+10.9$ \kms\ in the direction of \vz.  For comparison, the
  Mihalas \& Binney (1981) definition of the LSR requires a shift of
  the velocities used here by $-1.16$ \kms.}  When used as a model
optical depth distribution and converted to an absorption line
profile, the \HI\ emission data are a reasonable approximation of the
\ari\ profiles.  The other channels were then shifted to match the
LiF1A LSR velocities.  We estimate that our approach to determining
the absolute velocity zero point for the \fuse\ observations is
accurate to $\sim \pm5$ \kms. 

\begin{figure}
\epsscale{0.9}
\plotone{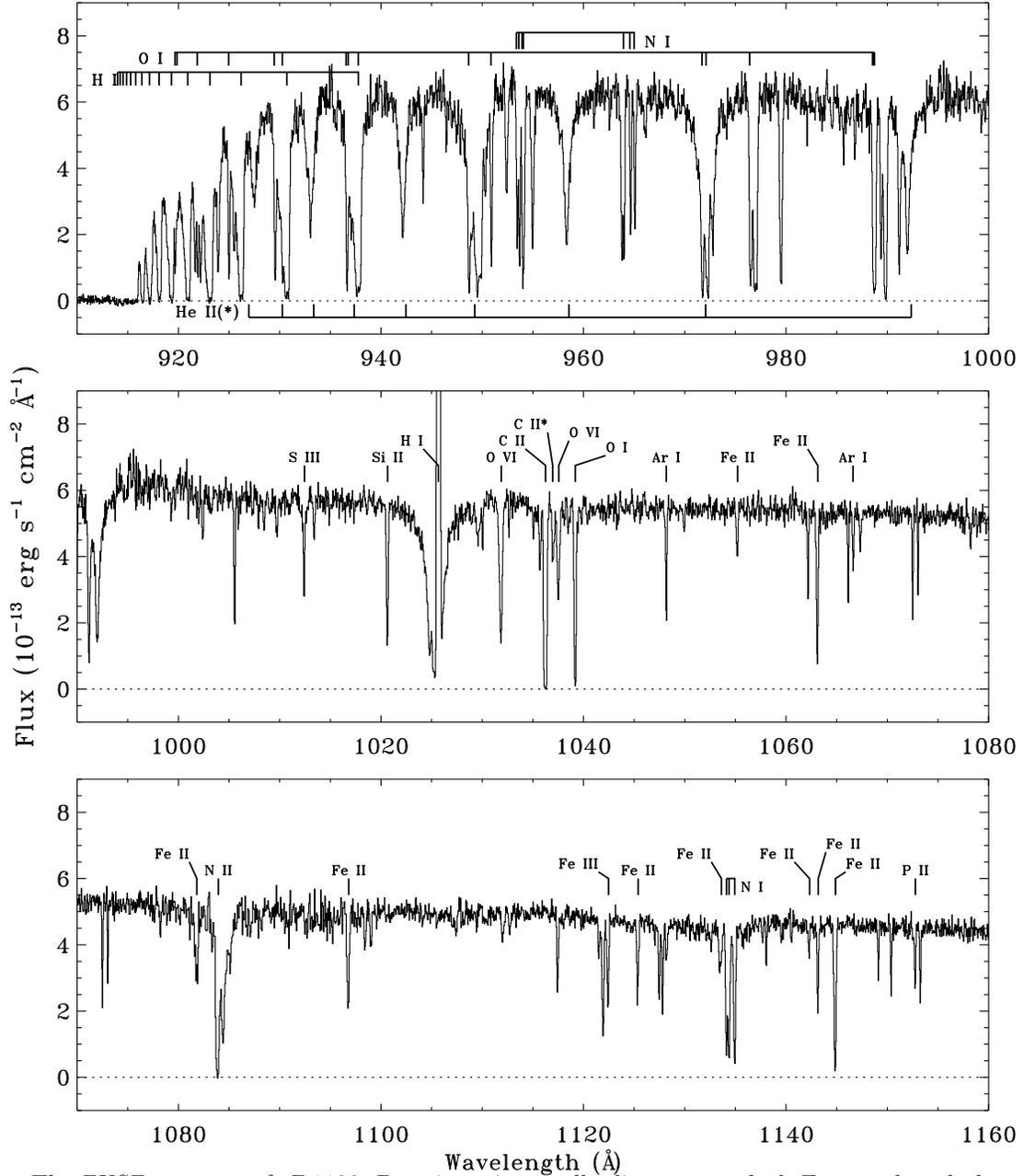}
\caption{The \fuse\ spectrum of vZ 1128.  Prominent interstellar lines
  are marked.  For wavelengths longward of 1000 \AA, the unmarked
  lines are all either stellar or blends of stellar and interstellar
  features.  Shortward of 1000 \AA\ we have only marked interstellar
  transitions of \HI, \ion{O}{1}, and \ion{N}{1}.  We have also noted
  the positions of stellar \ion{He}{2} transitions in this wavelength
  range below the spectrum in the top panel.
\label{fig:fullspec}}
\end{figure}

Figure \ref{fig:fullspec} shows the full \fuse\ spectrum of \vz, with
prominent interstellar features marked.  The displayed data are taken
from several different detector channels, with the majority from
SiC1B, LiF1A, and LiF2A.  Most of the unmarked absorption lines in
Figure \ref{fig:fullspec} are narrow stellar features.  We have
measured the centroids of several (10) of these (Table
\ref{tab:stellarvelocities}) and determine an average velocity of
$v_{\rm LSR} = -140\pm8$ \kms.  Our determination compares favorably
with the determination of Strom \& Strom (1970) who find $v_{\rm LSR}
= -142\pm15$ \kms.  Our quoted uncertainty includes contributions from
the uncertainties in the absolute velocity calibration and the \fuse\ 
relative wavelength scale.  However, the standard deviation in our
individual measurements, $\sigma = \pm2.4$ \kms, is significantly
smaller than the $\pm6$ \kms\ relative wavelength calibration
uncertainties typically quoted for \fuse\ data.  Due to the nature of
the \fuse\ detectors (see Sahnow et al. 2000), certain regions of the
detector (certain wavelengths) are better fit by the average
dispersion solution than others.  Thus, the $\pm6$ \kms\ uncertainty
quoted is a worst-case value.  We note that consistent measurements of
stellar velocities in wavelength regions near the \ovi\ 1031.926 \AA\ 
and \ion{Ar}{1} 1066.660 \AA\ transitions imply the relative
velocities of these two interstellar absorption lines are very well
calibrated (to within a few \kms).

Figure \ref{fig:fullspec} demonstrates an important benefit of using
hot post-AGB stars to probe the ISM: the stellar continuum is
extremely simple.  In large part this is because post-AGB stars
generally lack the prominent stellar P-Cygni profiles caused by the
line-driven winds of most early-type stars.  Furthermore, while many
photospheric absorption lines are present, they are quite narrow (with
breadths smaller then the ISM lines for \vz) and at velocities that
generally place them far from interstellar lines of interest (at least
for \vz).  Indeed, the spectrum of \vz\ is comparable in simplicity to
many AGN spectra but is at least twice as bright as 3C273 (e.g.,
Sembach et al. 2001).

\section{Analysis of Interstellar Absorption Lines}
\label{sec:analysis}

\begin{figure}
\epsscale{0.75} 
\plotone{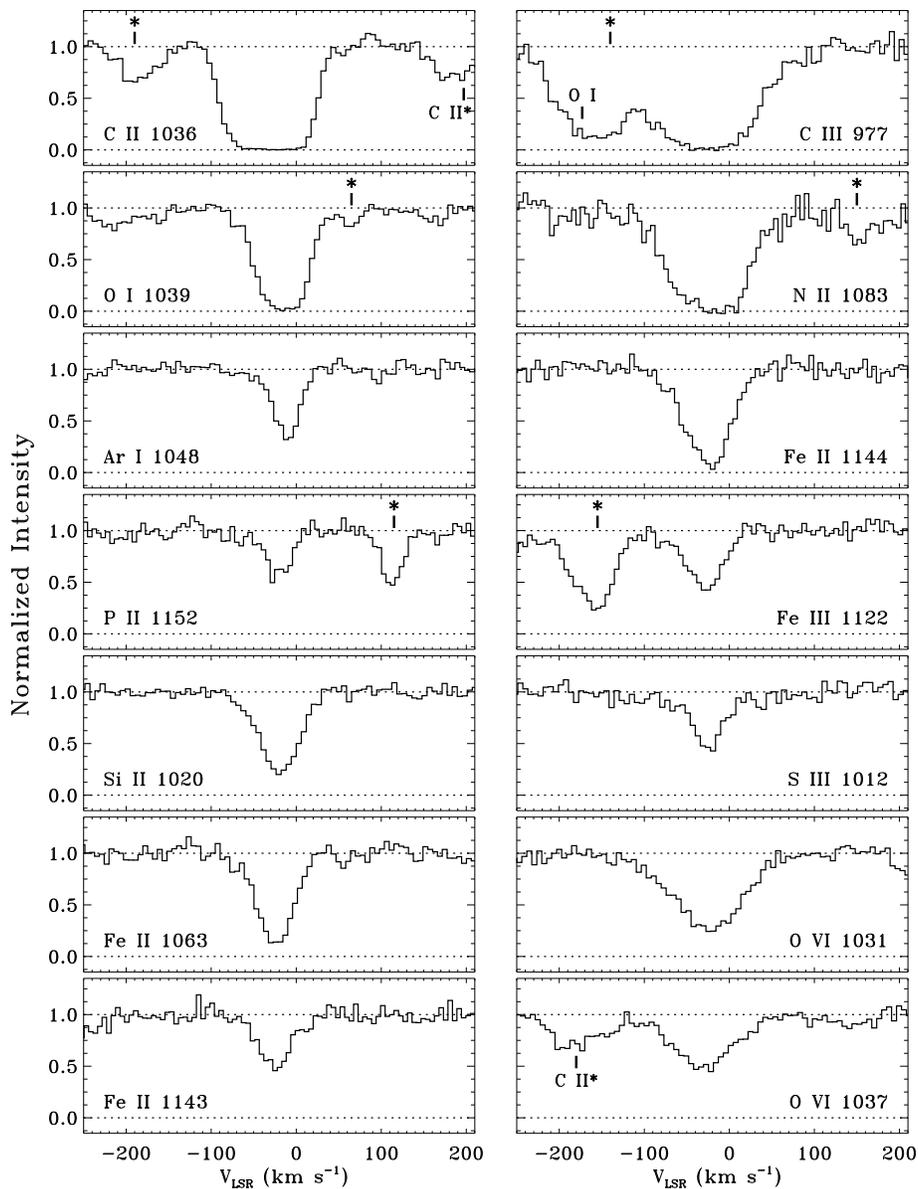}
\caption{Normalized profiles of several interstellar transitions versus
  LSR velocity, the majority taken from \fuse\ LiF1A and LiF1B
  spectra.  These data have been binned by three detector pixels to
  20.1 m\AA, or $\sim6$ \kms, per data point.  Stellar features and
  stellar/interstellar blends are marked with an asterisk.  Other
  interstellar transitions in the velocity range are marked.  The
  \ion{C}{3} and \ion{N}{2} profiles shown are from the SiC1B and
  SiC1A detector segments, respectively.  Each of these profiles has
  significant contamination from stellar absorption.  We have used a
  stellar model to remove most of the overlapping stellar absorption
  from the \ion{N}{2} profile (see text).
\label{fig:stack}}
\end{figure}

Each interstellar feature was normalized by fitting a low-order
Legendre polynomial to the adjacent stellar continuum.  For \vz, this
fitting process was for the most part unambiguous due to the lack of
interfering stellar features (the features that are present are easy
to distinguish and avoid in the fitting process).  The normalized
absorption profiles for several important interstellar transitions are
shown in Figure \ref{fig:stack}.  The empirically-determined
signal-to-noise ratios of these data are between $\sim15$ and 25 per
20 \kms\ resolution element.

We measured equivalent widths of interstellar transitions following
Sembach \& Savage (1992), including their treatment of the
uncertainties.  Equivalent width measurements are given in Table
\ref{tab:measurements} for numerous interstellar transitions present
in the \fuse\ data.  The uncertainties contain both statistical
(photon and fixed-pattern noise) and continuum placement
uncertainties.  For most transitions we quote equivalent widths
measured in detector 1 and 2 data separately.  For wavelengths
$\lambda > 1000$ \AA\ the measurements are from LiF1 and LiF2 data,
for $\lambda < 1000$ \AA\ from SiC1 and SiC2 data.  The velocity
ranges over which the equivalent width integrations were carried out
are given in the last column of Table \ref{tab:measurements}.  The
adopted atomic data are also summarized in this table.  The rest
wavelengths and most of the $f$-values are drawn from the revised
compilation of D. Morton (1999, private communication), except where
noted below.

Table \ref{tab:measurements} also gives the integrated apparent column
densities (Savage \& Sembach 1991) for each of the transitions using
the atomic data given in the second and third columns.  The apparent
optical depth, $\tau_a(v)$, is an instrumentally-blurred version of
the true optical depth of an absorption line, given by
\begin{equation}
  \tau_a(v) = - \ln \left[ I(v)/I_c (v) \right]
  \label{eqn:tauv}
\end{equation}
where $I_c(v)$ is the estimated continuum intensity and $I(v)$ is the
observed intensity of the line as a function of velocity.  This is
related to the apparent column density per unit velocity, $N_a(v)$ [${
  \rm atoms \ cm^{-2} \ (km \ s^{-1})^{-1}}$], by
\begin{equation}
  N_a(v) = \frac{m_e c}{\pi e^2} \frac{\tau_a (v)}{f \lambda} = 3.768
  \times 10^{14} \frac{\tau_a (v)}{f \lambda}, 
\end{equation}
where $\lambda$ is the wavelength in \AA, and $f$ is the atomic
oscillator strength.  In the absence of {\em unresolved} saturated
structure the $N_a(v)$ profile of a line is a valid,
instrumentally-blurred representation of the true column density
distribution as a function of velocity, $N(v)$.  In the presence of
unresolved saturated structure, the $N_a(v)$ distribution is always a
lower limit to the true column density distribution.

Table \ref{tab:columns} gives our adopted interstellar column
densities for the \vz\ sight line.  Two methods are used to determine
these values and limits.  For cases where it is expected that the line
profiles are fully resolved or that no saturation corrections need be
made, we make use of the apparent column density measurements from
Table \ref{tab:measurements} to determine the column densities.  In
the case of interstellar \ovi\ absorption, the absorption profiles are
typically broad enough to be fully resolved by the \fuse\ 
spectrographs.  For the \vz\ sight line, this is demonstrated by the
fact that the weak and strong members of the doublet give consistent
apparent column densities (see Table \ref{tab:measurements}).  For
other species, this is not necessarily the case.  In cases where no
saturation correction is required we adopt the straight average of the
apparent column density measurements.  In these cases the errors were
determined by assuming that the continuum placement uncertainties are
not lowered by averaging the individual measurements (we adopt the
minimum), while statistical uncertainties are lowered by $\sqrt{\sum
  \sigma_i^2} / N$, where $\sigma_i$ are the individual uncertainties
and $N$ the number of lines averaged.  The continuum and statistical
uncertainties are then added in quadrature.

We also use the apparent column density integrations from Table
\ref{tab:measurements} for deriving lower limits where saturation
corrections seem untenable (i.e., for strongly saturated lines such as
\cii\ $\lambda1036$).  In these cases we adopt the highest (most
constraining) lower limits available.  We derive upper limits for
undetected species assuming they are distributed as a Gaussian with
breadths equal to those observed for other species (see \S
\ref{subsec:osixkinematics} and \S \ref{subsec:kinematics}).  All
quoted upper limits have a $3\sigma$ confidence.  We note that no
molecular hydrogen is seen along this sight line; our limit on the
\htwo\ column density is $\log N(\mbox{\htwo}) \la 14.35$ ($3\sigma$),
summing all rotational states $J\leq3$.  The absence of \htwo\ implies
the line of sight does not contain significant cold neutral gas.

For species where saturation corrections could be important, we derive
column densities by fitting a single component, Doppler-broadened
curve of growth to the measured equivalent widths (fitting
measurements from both detectors simultaneously) or by applying the
curve of growth derived for similar species.  We have used
curve-of-growth fitting (in a manner following Howk et al. 2000) for
determining column densities of \ari\ and \feii\ using all of the
available measurements.  We have used the resulting $b$-values to
provide further information on the degree of saturation expected for
\ion{P}{2}, \sthree, and \fethree.  A few notes about specific species
are in order:

{\em C$\;${\small II} -- } The profile of the strong \ion{C}{2}
1036.337 \AA\ transition shows the entire velocity extent of the gas
along this sight line.  We have not presented measurements of
\ion{C}{2}$^*$ at 1037.018 \AA\ for this sight line.  The
\ion{C}{2}$^*$ absorption (which can be seen in the panel showing
\ion{O}{6} $\lambda1037.617$ in Figure \ref{fig:stack}) is
contaminated by unidentified stellar features.  While this could yield
an upper limit to the \ion{C}{2}$^*$ column density, we have not used
the profile for this purpose given the possibility that some
saturation exists in the profile (which would keep us from deriving a
true upper limit).

{\em P$\;${\small II} -- } The LiF1B and LiF2A measurements of
\ion{P}{2} are inconsistent with one another at the $3\sigma$ level.
The origins of this discrepancy are not clear, although the LiF1B
velocity profile seems inconsistent with those seen in other species.
We proceed by adopting the value derived using LiF2A, but given the
discrepancies between the two channels, we will make little use of the
\ion{P}{2} measurements.  Adopting the LiF2A measurements, we have
derived curve-of-growth column densities assuming
$b(\mbox{\ion{P}{2}}) \ga b(\mbox{\ari})$.  The inequality is adopted
because of the belief that \ari\ is more likely to reside in cooler
regions of the ISM. However, even adopting the relatively small
(compared with \feii) \ari\ \bval\ gives results that are consistent
(within $\sim1\sigma$) with the apparent column density
determinations.  For this reason we adopt the average of the apparent
column density and curve of growth methods.

{\em Fe$\;${\small II} -- } For \feii\ we have adopted the $f$-values
from Howk et al. (2000) with the exception of that for the transition
at 1144.938 \AA, for which we have adopted the new laboratory
measurement by Wiese, Bonvallet, and Lawler (2002).  However, for
comparison we have also fit a curve of growth adopting the FUV \feii\ 
oscillator strengths derived by Mallouris et al. (2001).  We find the
column densities and $b$-values derived with these two sets of
oscillator strengths to be consistent within the $1\sigma$
uncertainties.

{\em Fe$\;${\small III} --} We have derived \fethree\ column densities
assuming this ion has a velocity distribution similar to that of
\feii, adopting $b(\mbox{\fethree}) = 22.9\pm2.0$ \kms.  That is, we
apply the \feii\ curve of growth (with \bval\ uncertainties twice that
of \feii) to the \fethree\ equivalent width measurements.  We derive
column densities in this way that are indistinguishable from those
derived using the apparent column density method.

{\em S$\;${\small III} --} We assume that this ion is distributed in a
manner like \fethree, which implies no saturation correction.  This is
justified by the similarity of the \sthree\ and \fethree\ profiles
(see \S \ref{subsec:kinematics}).

{\em N$\;${\small II} --} Though this line only yields a lower limit
to the \ntwo\ column density, the analysis that led to this limit is
sufficiently different than that used for the other species that it
requires description.  The strong broad stellar \ion{He}{2} $1084.940$
\AA\ transition overlaps the positive velocity wing of the \ntwo\ 
1083.994 \AA\ interstellar transition (the center of the \ion{He}{2}
is found at $\sim+120$ \kms\ relative to the restframe of \ntwo,
including the effects of the large stellar velocity).  Weak absorption
from stellar \ntwo$^*$ $\lambda\lambda 1084.566, 1084.584$ is also
present at lower positive velocities ($\sim+20$ \kms) in the restframe
of the ground state interstellar transition.  We have accounted for
the overlap of these stellar features with the interstellar absorption
by using a model atmosphere provided by P. Chayer (2001, private
communication) to normalize the data near \ntwo\ $\lambda1083.994$
before processing the data as described above.  The model atmosphere
was calculated in a manner similar to that described in Sonneborn et
al. (2002) using the TLUSTY code of Hubeny \& Lanz (1995) and the
SYNSPEC code of I. Hubeny (2000, private communication).  The model is
characterized by $T=35,000$ and $\log g = 4.0$ (Dixon et al. 1994).
The model assumes a ratio ${\rm He/H}=0.1$ and a metallicity $[{\rm
  M/H}]=-2$.

\section{Highly-Ionized Gas Toward von Zeipel 1128}
\label{sec:osix}

The hot phase of the ISM in galaxies is a direct result of the
feedback of energy and matter into the ISM from early-type stars.  The
hot ($T\ga10^6$ K) ionized gas in the Galactic ISM is typically traced
through its X-ray emission, particularly its soft X-ray emission.
Much of this hot material will eventually cool through various
processes, either through simple radiative cooling (e.g., as
envisioned in galactic fountain scenarios; Edgar \& Chevalier 1986) or
through more complex mechanisms where the hot gas interacts with
cooler material (e.g., through conduction or mixing; see review by
Spitzer 1996).  Material cooling from X-ray emitting temperatures
passes very quickly through the temperature regime of a
few$\times10^5$ K due to the presence of strong emission lines of
highly-ionized metals in this temperature range, especially the
lithium-like ions of carbon, nitrogen, and oxygen.  For this reason,
gas in this regime is often referred to as ``transition temperature''
gas or, by analogy with recent work on the intergalactic medium,
warm-hot ionized gas.

It is this transition temperature gas that is traced by \ovi\ 
absorption toward \vz\ and other background ultraviolet continuum
sources.  This section discusses the properties of the highly-ionized
gas in the Galactic thick disk toward \vz.  While \ovi\ absorption
line measurements provide good kinematic information on the warm-hot
ionized medium, it should be emphasized that this material does not
represent the majority of the hot ionized gas in the thick disk.
Instead, \ovi\ is a tracer of the cooling of the hot ionized medium or
of its interaction with cooler material.

\subsection{Column Densities of Highly-Ionized Gas}

The sight line to \vz\ probes the entire interstellar thick disk of
the Milky Way and parts of the more extended corona.  The distribution
of \ovi\ in the former region is best described with an exponential
scale height of $h_z \sim 2.3$ kpc having an excess in the northern
polar cap region of $\sim0.25$ dex (Savage et al.  2003; Zsarg\'{o} et
al. 2003).  The midplane density of the distribution is
$n_0(\mbox{\ovi}) \approx 1.7\times10^{-8}$ \percc\ (Jenkins, Bowen,
\& Sembach 2001).  These quantities suggest an average value $\log
N(\mbox{\ovi}) \sin |b|$ of $14.08+0.25=14.33$ integrated through the
northern extent of the Galaxy.  It should be noted that the \ovi\ 
layer is very patchy (Howk et al. 2002b; Savage et al. 2003), with a
standard deviation about the mean $N(\mbox{\ovi}) \sin |b|$ of
$\sim0.19$ dex ($\sim40\%$) for sight lines probing the entire
Galactic halo.

The \ovi\ column density observed toward \vz, $\log N(\mbox{\ovi})
\sin |b| \approx \log N(\mbox{\ovi}) = 14.49\pm0.03$, is larger than
the northern Galactic average value determined by Savage et al. (2003)
in their survey of \ovi\ toward extragalactic continuum sources.
However, given the large observed dispersion in the distribution of
\ovi\ column densities, that toward \vz\ does not
seem unusually large.  We note that the great distance of \vz\ makes
the column densities derived here more comparable to the extragalactic
sample of targets than, for example, to the halo star sample studied
by Zsarg\'{o} et al. (2003).

In principle the column density ratios of various highly-ionized
species can be used to constrain the temperature of the gas, if
collisional ionization dominates (as we expect for \ovi).  For this
reason we have searched for \ion{S}{6} along this sight line, which
has creation and destruction ionization potentials of 72.7 and 88.0
eV.  While the \ovi\ column density is relatively large, there is no
evidence for absorption by interstellar \ion{S}{6} in the \fuse\ 
spectrum of \vz.  Assuming \ion{S}{6} to have the same velocity
dispersion as \ovi, we derive a $3\sigma$ limit of $\log
N(\mbox{\ion{S}{6}})/ N(\mbox{\ovi}) < -1.1$.  In the collisional
ionization equilibrium calculations of solar-metallicity gas by
Sutherland \& Dopita (1993), this ratio implies $T \ga 2\times10^5$ K.
We caution that equilibrium conditions may not hold. For comparison,
the non-equilibrium calculations of Shapiro \& Moore (1976), in which
collisionally-ionized gas at $T\sim10^6$ K is allowed to cool
isochorically to $10^4$ K, are consistent with this ratio for $10^4 <
T \la 6\times10^5$ K.  Evidently the ratio
$N(\mbox{\ion{S}{6}})/N(\mbox{\ovi})$ does not by itself place strong
constraints on the temperature of highly-ionized gas given the
discrepancies between the equilibrium and non-equilibrium
calculations.

Similarly, our data limit the ratio $\log N(\mbox{\ion{S}{4}})/
N(\mbox{\ovi}) < -0.8$ ($3\sigma$).  This ratio is also consistent
with the Sutherland \& Dopita (1993) equilibrium models for $T\ga
2\times10^5$ K.  The Shapiro \& Moore (1976) calculations predict
values in agreement with our limit for $T\ga 1.5\times10^5$ K;
however, their calculations for $1.2\times10^4 \la T \la
7.2\times10^4$ K are also consistent with our limits on
$N(\mbox{\ion{S}{4}})/ N(\mbox{\ovi})$.  In this case the temperature
limits derived from the equilibrium and non-equilibrium models overlap
better than for the $N(\mbox{\ion{S}{6}})/N(\mbox{\ovi})$ ratio.
However, one must still be concerned that these models are not the
best descriptor of the situation at hand.  (It is also important to
note that some \ion{S}{4} could arise in photoionized material,
similar to the case for the well-studied ion \ion{Si}{4}.)

We have also placed limits on the column density of \ion{P}{5} (with
creation and destruction ionization potentials 51.4 and 65.0 eV)
toward \vz, deriving $\log N(\mbox{\ion{P}{5}})/N(\mbox{\ovi}) < -1.8$
($3\sigma$).  Neither Sutherland \& Dopita nor Shapiro \& Moore report
the ionization balance of phosphorus.  However, with $\log ({\rm
  P/O})_\odot = -3.29$ (Grevesse \& Sauval 1998), this limit is not
likely to be useful.

We can estimate the column density of protons associated with the hot
ISM in this direction using the observed \ovi\ column density.  The
proton column density associated with a tracer of ionized gas, $X^i$,
is:
\begin{equation}
  N(\mbox{\HII}) \approx N(X^i) \, ({\rm  X/H})^{-1} \, x(X^i)^{-1},
\label{eqn:protoncolumn}
\end{equation}
where $N(X^i)$ is the column density of the ion $X^i$, $({\rm X/H})$
is the abundance of X/H (usually assumed to be equivalent to the solar
system abundance), and $x(X^i) \equiv n(X^i)/n(X)$ is the ionization
fraction of the ion $X^i$.  In the case at hand, we have
$N(\mbox{\HII}) \approx N(\mbox{\ovi}) \, ({\rm O/H})^{-1} \,
x(\mbox{\ovi})^{-1}$.  We assume the abundance of oxygen, O/H, to be
the solar value of $\log \, ({\rm O/H})_\odot = -3.26$ (Holweger 2001)
and that the fraction of oxygen in the form of \ovi, $x(\mbox{\ovi})$,
is $\la0.2$, which is consistent with both the equilibrium
calculations of Sutherland \& Dopita (1993) and the non-equilibrium
calculations of Shapiro \& Moore (1976).  With these assumptions we
derive $\log N(\mbox{\HII}) \ga 18.4$ for the transition-temperature
($\sim10^5-10^6$ K) gas toward \vz.

\subsection{Kinematics of the Highly-Ionized Gas}
\label{subsec:osixkinematics}

\begin{figure}
\epsscale{0.9}
\plotone{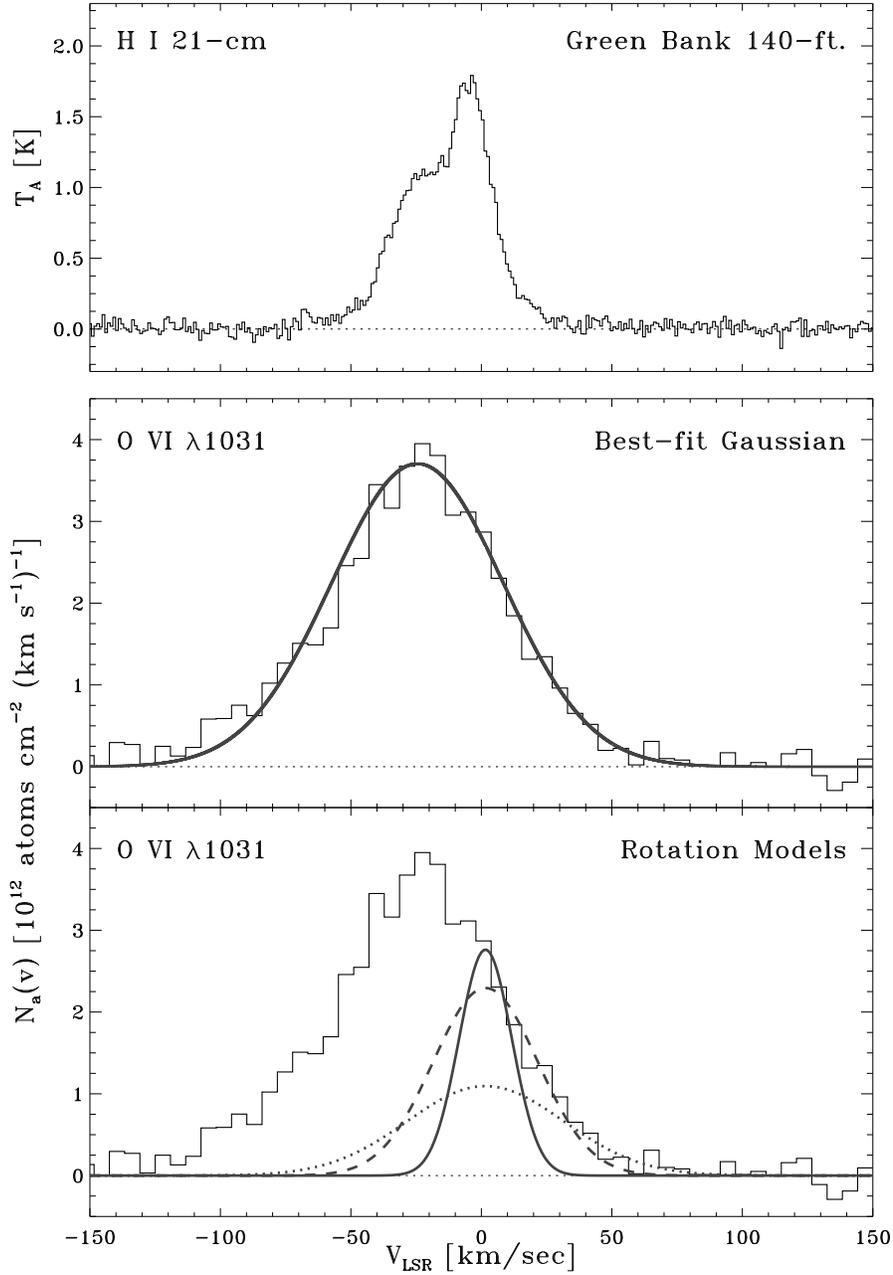}
\caption{The Green Bank 140-ft \ion{H}{1} profile ({\em top}) and the 
  \ion{O}{6} apparent column density profile ({\em bottom two panels})
  observed toward vZ~1128.  The \ion{O}{6} $N_a(v)$ profiles are shown
  with the best-fit Gaussian overlaid ({\em middle}) and three models
  of the $N_a(v)$ distribution assuming the \ion{O}{6} participates in
  Galactic rotation ({\em bottom}).  The best-fit Gaussian has a
  dispersion $\sigma = 31.8\pm1.1$ (after removal of instrumental
  effects).  The rotation models all assume the \ion{O}{6} is
  distributed with a scale height of 2.3 kpc.  The three models differ
  in their assumed cloud velocity dispersion (using $\sigma = 10$, 20,
  and 30 km s$^{-1}$) and have been scaled to best match the red wing
  of the \ion{O}{6} profile. \label{fig:oviprofile}}
\end{figure}

Figure \ref{fig:oviprofile} shows the \nav\ distribution of \ovi\ 
toward \vz\ and the NRAO \HI\ 21-cm spectrum.  The middle panel shows
the \ovi\ \nav\ profile with the best-fit Gaussian distribution
overlayed.  The properties of this Gaussian are summarized in Table
\ref{tab:kinematics}, along with the properties of Gaussians fit to
the \nav\ profiles of several lower-ionization species.  The best-fit
Gaussian dispersion, $\sigma \, (\equiv b/\sqrt{2})$, after removal of
the instrumental breadth, is $31.8\pm1.1$ \kms\ (equivalent to a
Doppler parameter $b = 45.0$ \kms).  This limits the temperature of
the \ovi\ along the sight line to $T\la 2\times10^6$ K.  This is
significantly larger than the temperature at which the abundance of
\ovi\ peaks in collisional ionization equilibrium, $T\sim3\times10^5$
K, and the large breadth indicates significant non-thermal broadening
of the profile.

The kinematic properties of the \ovi\ profile toward \vz\ are
significantly different than expected for a highly-ionized thick disk
corotating with the underlying thin disk.  Because of the high
latitude of this star, Galactic rotation is expected to have very
little impact on the velocity profiles of any interstellar species.
The bottom panel of Figure \ref{fig:oviprofile} shows the \ovi\ \nav\ 
profile with three models of simple rotation of a thick \ovi -bearing
disk with scale height of $\sim2.3$ kpc (see Savage et al. 2003).  The
three models differ in the assumed dispersion of the \ovi -bearing gas
(with models shown for Gaussian dispersion of $\sigma = 10$, 20, and
30 \kms).  The midplane density in each model is scaled to match the
red wing of the \ovi\ profile.

The observed velocity distribution of \ovi\ is clearly inconsistent
with these simple models of Galactic rotation.  The observed profile
shows much more material at negative velocities than is expected for a
corotating thick disk of \ovi -bearing material.  Indeed, the velocity
centroid of \ovi\ suggests the bulk of the highly-ionized material
along this sight line is falling onto the Galactic disk from the halo.
The central velocity of the \ovi\ profile, $\langle \mvlsr \rangle
\approx -26$ \kms, is consistent with the center of the broad negative
velocity absorption seen in the NRAO \HI\ profile (at $\mvlsr \sim
-25$ \kms).  It is also interesting to note the relatively good
agreement between the central velocities of WNM/WIM tracers and \ovi\ 
(see Table \ref{tab:kinematics}).  Thus, it would seem that the
infalling gas along this sight line contains a range of ionization
states.  We discuss the kinematic relationship between the WNM/WIM
tracers in \S \ref{subsec:kinematics}.

The kinematics of \ovi\ along the \vz\ sight line are perhaps best
discussed in the context of similar high-latitude observations.
Wakker et al. (2003) and Savage et al. (2003) present Galactic \ovi\ 
absorption lines toward $\sim100$ extragalactic objects. They find
that while there is a large spread in the observed central velocities
of the \ovi\ observed along high-latitude sight lines, the average of
the distribution is $\mvlsr = 0\pm21$ \kms\ (standard deviation).
Thus, while we observe the \ovi\ to be infalling toward \vz, this
should be viewed in the context of the broad distribution of
outflowing and infalling \ovi\ gas across the entire high-latitude
sky.  The central velocity of the \vz\ \ovi\ distribution cannot be
used, for example, to claim that the entire highly-ionized Galactic
thick disk is collapsing onto the plane.

The observed Gaussian dispersion can also be discussed in the broader
context of the Wakker et al. and Savage et al. measurements.  These
authors use a slightly different approach to determining the velocity
breadths of the \ovi\ absorption profiles, defining a quantity we
refer to as $b^\prime$:
\begin{displaymath}
   b^\prime \equiv 
      \left[ 2 \, \frac{\int (v-\bar{v})^2 N_a(v) dv}
                    {\int N_a(v) dv} \right]^{1/2}
\end{displaymath}
These authors find $\langle b^\prime \rangle = 60\pm15$ \kms\ for
their sample of extragalactic sight lines (with no correction for
instrumental effects).  Our measured dispersion corresponds to $b
\approx 46.5\pm1.6$ \kms\ (before correcting for the instrumental
spread function).  If derived in the manner used by these authors, we
find $b^\prime$ for the \ovi\ toward \vz\ is $57\pm3$ \kms.  Thus, the
breadth of the \ovi\ profile toward \vz\ is well within one standard
deviation of the high-latitude mean derived by Wakker et al.  and
Savage et al.



\section{Warm Ionized and Neutral Gas Toward von Zeipel 1128}
\label{sec:warm}

One of the goals of this program is to understand the relationship
between the observational tracers of warm neutral and ionized gas in
the Galactic thick disk.  This section discusses the relative
elemental abundances in the WNM and WIM as well as the kinematics of
the tracers of these phases toward \vz.

The WNM and WIM in the Galactic thick disk are probed by low and
intermediate ions, respectively.  Low-ionization species include those
thought to be the dominant ionization state of each element in the
WNM; low ions observed toward \vz\ include \cii, \ion{N}{1}, \oi,
\ion{Si}{2}, \ion{P}{2}, \ari, and \feii.  We describe the next-higher
ionization stages as intermediate ions, including the species \cthree,
\ntwo, \sthree, and \fethree\ observed toward \vz.  These ions are
likely to arise in the WIM (and perhaps to some degree in the hot ISM;
see Howk \& Savage 1999b) along with some of the low ions.

The low ions may arise in both the WNM and WIM (Haffner, Reynolds, \&
Tufte 1999; Sembach et al. 2000), which is a source of some concern,
since these are the ions used to determine gas-phase abundances in the
WNM (see discussion in Sembach et al. 2000).  The degree of
uncertainty associated with this ``contamination'' (or confusion) of
WNM species by gas associated with the WIM varies from element to
element.  We discuss such contamination for the specific case of iron
below.

\subsection{Gas-phase Abundances in the Warm Neutral Material}
\label{sec:abundances}

The gas-phase abundance of an element $X$ relative to an element $Y$
is calculated as
\begin{equation}
  [X/Y] \equiv \log \, [N(X^i)/N(Y^j)]
  - \log \, (X/Y)_\odot - \log \, [x(X^i)/x(Y^j)] .
\label{eqn:abundance}
\end{equation}
In this expression $N(X^i)$ and $N(Y^j)$ are the column densities of
the ions $X^i$ and $Y^j$.  These column densities are normalized to
the relative solar system abundances of the two elements,
$(X/Y)_\odot$ (Grevesse \& Sauval 1998; see Table
\ref{tab:wnmabundances}).  The quantities $x(X^i)$ and $x(Y^j)$ are
the ionization fractions of each ion in the gas being studied.  The
last term is the ionization correction.  Typically the dominant WNM
ionization state of the two elements is used, and the ionization
correction is assumed to be insignificant.  Exceptions to this general
rule include studies of \ari, which is easily photoionized in WNM
conditions (Sofia \& Jenkins 1998).

\begin{figure}
\epsscale{0.9}
\plotone{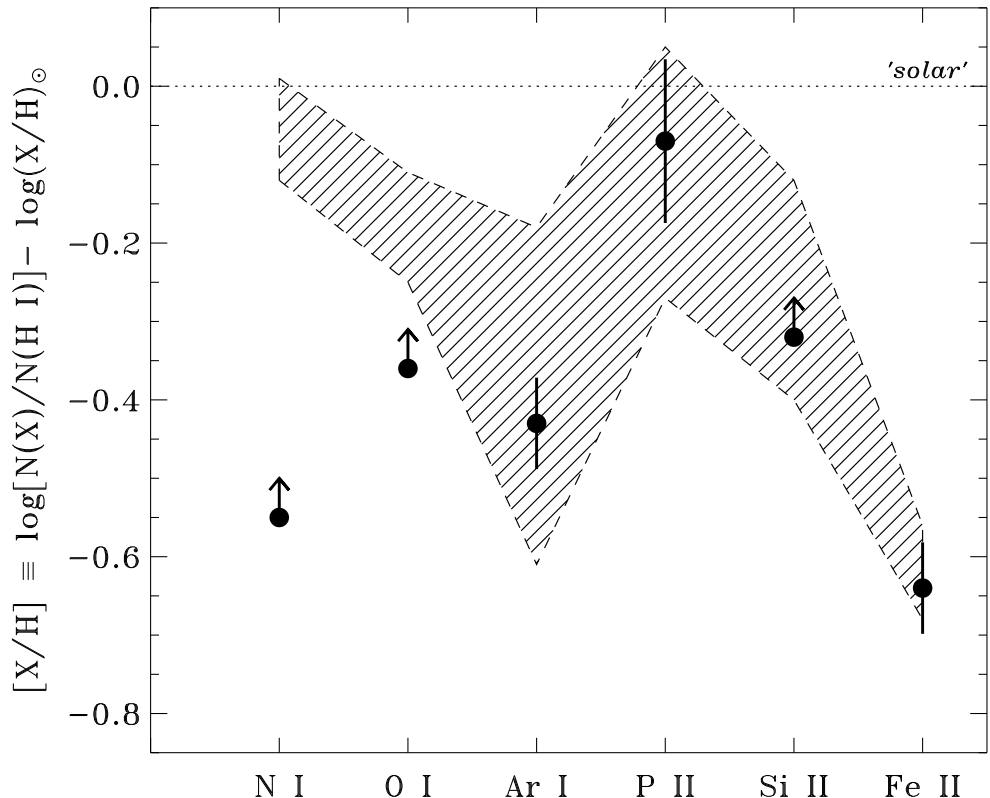}
\caption{Normalized gas-phase abundances of species found in the 
  warm neutral medium along the vZ 1128 sight line ({\em points}).  No
  corrections have been made for ionization effects.  The hatched area
  represents the range of values commonly found along Galactic halo
  sight lines for silicon and iron.  For nitrogen and oxygen, we use
  the $1\sigma$ confidence intervals of the distribution of measured
  values from Meyer et al. (1997) and Andre et al. (2003),
  respectively.  For argon the values represent the range measured by
  Sofia \& Jenkins (1998).  The values given for phosphorous, which
  has few measurements for low density environments, represent the
  full range of recent measurements: the lower point from Welty et al.
  (1999) and the upper point from Howk et al. (1999).  We note that
  Jenkins, Savage, \& Spitzer (1986) derive an average value of $[{\rm
    P/H}] \approx -0.2$ for WNM sight lines studied with {\em
    Copernicus}. We assume solar system abundances from Grevesse \&
  Sauval (1998), except for argon (6.48; from Sofia \& Jenkins 1998)
  and nitrogen and oxygen (7.93 and 8.74, respectively; from Holweger
  2001). \label{fig:wnmabundances}}
\end{figure}

Table \ref{tab:wnmabundances} and Figure \ref{fig:wnmabundances}
summarize the gas-phase abundances in the WNM derived in the usual
way.  That is, we have compared the column densities of low-ionization
species with that of \HI\ in this direction.  Studies of the WNM
typically use \HI\ or a non-depleted metal ion as a reference for
studying interstellar grain composition.  The relative deficit of
metals in the ISM compared with the solar system is assumed to be
caused by the incorporation of these metals into dust grains (the
solid phase).  Applying the standard technique yields only an
approximate determination of $[X/{\rm H}]$ for the WNM in this
direction because of the assumption that $\log \,
[x(X^i)/x(\mbox{\HI})] = 0$ in Eqn. (\ref{eqn:abundance}).

Figure \ref{fig:wnmabundances} shows the WNM gas-phase abundances for
the \vz\ sight line along with the range of values seen along other
thick disk (or ``halo'') sight lines.  The gas-phase abundances toward
\vz\ are not out of the ordinary for high-latitude sight lines.  We
point out that these measurements represent averages over the entire
sight line.  There is evidence that many of the metals are distributed
differently in velocity space than the \HI, especially the species
heavily incorporated into dust grains (see below).

\subsection{Ionization Effects}
\label{sec:ionization}

The measurements given in Table \ref{tab:wnmabundances} provide a
baseline for determining the extent to which ionization effects
influence such abundance determinations and for comparing the
properties of the WNM and WIM along the \vz\ sight line.  We first
discuss the average ionization state of the gas along this sight line
because it bears on the gas-phase abundances derived above.

Using the Eqn. (\ref{eqn:protoncolumn}), it is possible to estimate
the column density of protons associated with the intermediate ions
toward \vz.  Such an estimate, combined with measures of
$N(\mbox{\HI})$, provides a rough measure of the total hydrogen column
density along the sight line.  For determining the proton column
density of the WIM, we make use of the \sthree\ column density
reported in Table \ref{tab:columns}.

The most basic assumption, i.e., that all of the WIM sulfur is in the
form of \sthree\ [$x(\mbox{\sthree})=1$] and sulfur is undepleted
(Howk, Savage, \& Fabian 1999), results in a lower limit to the proton
column associated with the WIM: $\log N(\mbox{\HII}) > 19.1$.  Sembach
et al.  (2000) have presented models of the ionization of the WIM
assuming ionization by early-type stars.  If we adopt the sulfur
ionization fractions from their composite standard model
[$x(\mbox{\sthree})\sim0.18$] for our calculation, we find $\log
N(\mbox{\HII}) = 19.9$.  For comparison, similar calculations
performed using the lower and upper limits provided by \ntwo\ and
\ion{P}{3} yield $18.6 < \log N(\mbox{\HII}) \la 20.8$, consistent
with the valued derived using \sthree.

The proton columns estimated using \sthree\ above imply a total
hydrogen column density of $\log N(\mbox{\HI +\HII}) = 20.0$ to 20.2
and $\langle x(\mbox{\HII}) \rangle \equiv N(\mbox{\HII})/N(\mbox{\HI
  +\HII}) \ga 0.12$ to $\sim0.46$, with the latter being the preferred
estimate.  These values are roughly consistent with the estimates of
$x(\mbox{\HII})$ derived from the \halpha\ intensity in this general
direction (\S \ref{sec:sightline}).

With this estimate of the total hydrogen column density of the WNM and
WIM (since the hot ISM likely contributes little, as discussed in \S
\ref{sec:osix}), we can further estimate the total gas-phase iron
abundance.  The total iron column density along this sight line is
$\log N({\rm Fe})_{total} \approx \log N(\mbox{\feii +\fethree}) =
14.95\pm0.04$, assuming higher ionization states represent a small
fraction of the total.  Thus, the gas-phase abundance of iron along
the sight line to \vz\ (averaged over all phases and regions) is
\begin{displaymath}
  -0.76 \la {\rm [Fe/H]}_{total} < -0.56.  
\end{displaymath} 
This can be compared to the value $[{\rm Fe/H}]_{WNM} = -0.64\pm0.06$
derived using \feii\ and \HI\ with no ionization correction.  The
ionization corrections for the derived WNM gas-phase abundance appear
to be $\la \pm0.1$ dex (averaged over the entire sight line).

\subsection{Gas-phase Abundances in the Warm Ionized Material}

It is also possible to estimate the gas-phase abundances of refractory
elements in the WIM by comparing the columns of
intermediate-ionization stages of depleted and non-depleted species
and applying an ionization correction.  Howk \& Savage (1999b) have
presented determinations of [Al/S] in the WIM using this approach,
arguing on the basis of the derived values that dust is present in the
ionized medium of the Galaxy.  This conclusion has been confirmed by
the detection of far-infrared emission from dust associated with the
WIM (Lagache et al. 2000).

Our observations of \fethree\ and \sthree\ provide a similar
opportunity to determine the relative gas-phase abundances of iron and
sulfur in the WIM.  We determine [Fe/S]$_{WIM}$ in the manner outlined
in Eqn. (\ref{eqn:abundance}):
\begin{equation}
  {\rm [Fe/S]}_{WIM} \equiv \log \, [N(\mbox{\fethree})/N(\mbox{\sthree})] 
       - \log \, ({\rm Fe/S})_\odot 
       - \log \, [x(\mbox{\fethree})/x(\mbox{\sthree})] . 
\end{equation}
In this case the ionization correction term, $\log
[x(\mbox{\fethree})/x(\mbox{\sthree})]$, is significant.

The composite standard WIM model (Sembach et al. 2000) has $\log
[x(\mbox{\fethree})/x(\mbox{\sthree})] = +0.45$, which yields
\begin{displaymath} 
  {\rm [Fe/S]}_{WIM} \sim -0.74 \pm 0.3        
\end{displaymath}    
for the sight line to \vz.  The error bar is an estimate of the
uncertainties in the ionization corrections due to the sensitivity of
$x(\mbox{\fethree}) / x(\mbox{\sthree})$ to the adopted effective
temperature of the ionizing source following Howk \& Savage (1999b).
This determination is in excellent agreement with the WNM
determination and the values of ${\rm [Fe/H]}_{total}$ and ${\rm
  [Fe/H]}_{WNM}$ derived above.  The implications of this similarity
are discussed more thoroughly in \S \ref{sec:discussion}.

\subsection{Kinematics of the Warm Ionized and Warm Neutral Gas}
\label{subsec:kinematics}

\begin{figure}
\epsscale{0.6}
\plotone{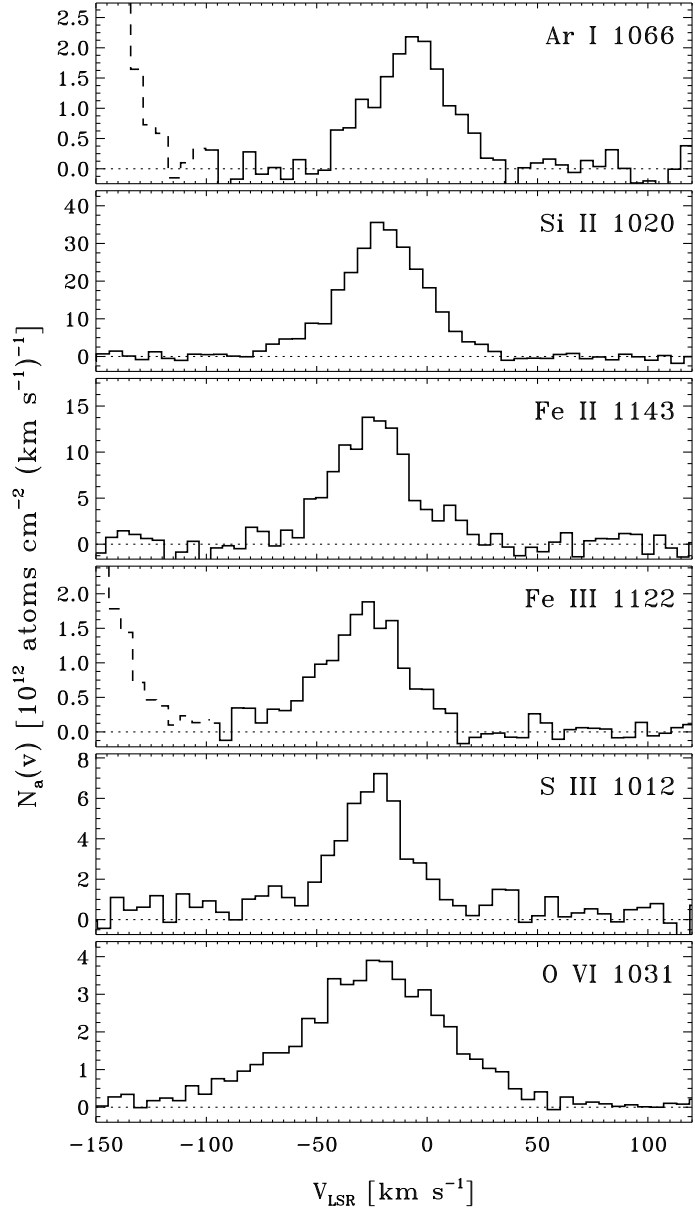}
\caption{Apparent column density profiles of several ionic species 
  observed toward vZ~1128 that trace the warm neutral (\ion{Ar}{1},
  \ion{Si}{2}, \ion{Fe}{2}), the warm ionized (\ion{Fe}{3},
  \ion{S}{3}), and the hot ionized (\ion{O}{6}) gas toward vZ~1128.
  The two singly-ionized species may also contain a contribution from
  the warm ionized medium.  It should be noted that the profiles of
  \ion{Ar}{1} and \ion{Si}{2} likely contain unresolved saturated
  structure that causes the $N_a(v)$ distribution to be different from
  the true column density distribution.  The analysis of the other
  transitions suggests relatively small saturation effects.  Dashed
  lines in several panels represent unrelated interloping absorption
  lines.
\label{fig:navstack}}
\end{figure}

We now turn our attention to the kinematic relationship between the
low- and intermediate-ions toward \vz.  Figure \ref{fig:navstack}
shows the apparent column density profiles of a range of ions observed
toward \vz, including species tracing the WNM and the WIM.  Also shown
is the \ovi\ profile, which traces the warm-hot ISM.  Measurements of
the velocity centroids and breadths of the \nav\ profiles of several
species are given in Table \ref{tab:kinematics}. The peak apparent
optical depths of each transition are also given as a guide for
judging the relative importance of saturation effects on the derived
kinematic properties.

It is clear that the resolution of \fuse\ is insufficient to clearly
resolve the detailed component structure of this sight line.
Examining the \nav\ profile of the strong \ion{Si}{2} 1020 \AA\ line
(which is susceptible to unresolved saturation at the highest optical
depths), there is clear evidence for velocity substructure that cannot
be clearly separated at \fuse\ resolution.  For this reason, we
concentrate our discussion of the kinematics of the WNM/WIM gas on the
central velocity and breadth of the absorption.  The majority of the
species presented in Figure \ref{fig:navstack} and Table
\ref{tab:kinematics} have similar central velocities ranging from
$\mvlsr \approx -31$ to $-21$ \kms, including \ovi.  The one exception
to this agreement in the velocities of the observed species is
\ion{Ar}{1}, which is centered near $\mvlsr\approx-10$ \kms.  Because
of its large ionization cross section (Sofia \& Jenkins 1998),
\ion{Ar}{1} is relatively easy to ionize in diffuse interstellar gas,
even though its ionization potential is larger than that of \HI.  This
is the most likely cause for the apparent deficit of \ion{Ar}{1} seen
in Figure \ref{fig:wnmabundances}, particularly since argon is
unlikely to bind to grains effectively (Sofia \& Jenkins 1998).  All
of the species displayed in Figure \ref{fig:navstack} show absorption
at the velocities at which \ion{Ar}{1} has a maximum optical depth.

\begin{figure}
\epsscale{0.6}
\plotone{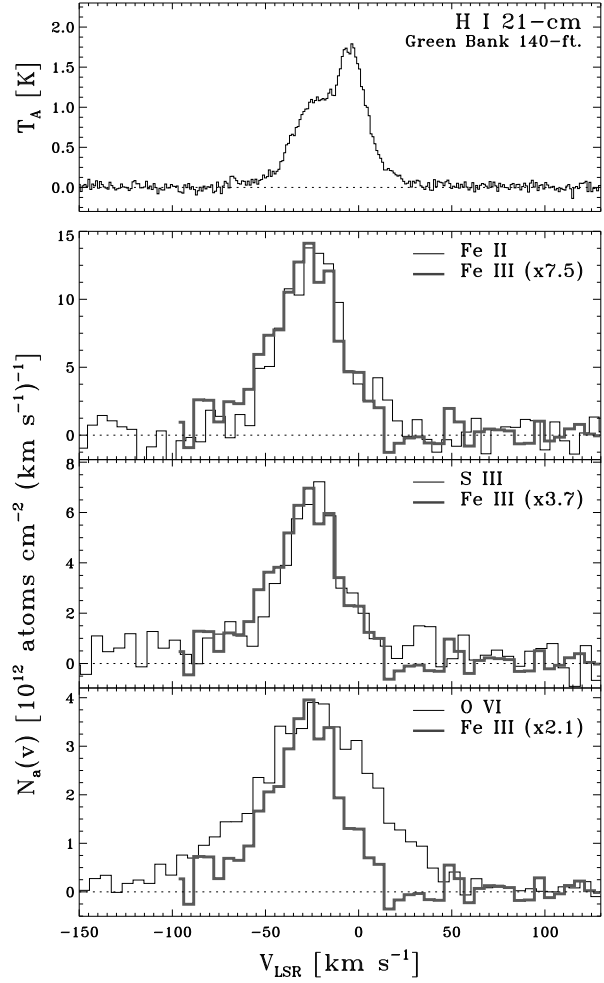}
\caption{Apparent column density profiles of \ion{Fe}{3} compared 
  with \ion{Fe}{2}, \ion{S}{3}, and \ion{O}{6}.  In each case the
  \ion{Fe}{3} profile has been scaled to match the other species.  The
  WNM and WIM species trace each other quite well (within the
  limitations of the {\em FUSE} resolution), though there may be a
  slight excess of \ion{Fe}{3} near $v_{\rm LSR} \approx -50$ km
  s$^{-1}$ compared with the others.  The \ion{Fe}{3} and \ion{O}{6}
  have nearly the same central velocity, though the latter, which
  traces hotter gas, is much broader.
\label{fig:wim}}
\end{figure}

To the extent that the \fuse\ resolution allows us to study such
coincidences, the breadths of the profiles of species tracing the WNM
and WIM are quite similar.  Figure \ref{fig:wim} compares the apparent
column density profiles for \ion{Fe}{2}, \fethree, and \sthree.  With
the exception of a possible weak excess in \fethree\ at $\mvlsr \sim
-50$ \kms, the profiles of these species track one another remarkably
well.  The similarity of the observed profiles suggests the WNM and
WIM share a similar component structure.  Large component-to-component
variations in the ratio of the WNM and WIM tracers can be ruled out
for the dominant component groups along this sight line (strong
variations in the ratio among low column density components cannot be
ruled out).

An interpretation of the kinematics of the WNM/WIM tracers discussed
in this section is that the species centered near $\mvlsr \sim -26$
\kms\ principally trace gas in the Galactic thick disk and are
associated with the broad \HI\ seen at similar velocities.  The
material closer to the LSR zero point, including the narrower peak of
the \HI\ distribution and the majority of the \ion{Ar}{1}, is likely
tracing gas associated with lower-ionization, higher-depletion
material in the Galactic disk in this direction.  The principal WNM
tracers accessible by \fuse\ probe elements that are readily
incorporated into dust grains.  These ions are therefore biased toward
gas in the Galactic thick disk, which is more likely to be ionized
than the thin disk and to have higher gas-phase abundances due to the
processing of dust grains (e.g., Sembach \& Savage 1996).

It is significant that there seems to be little evidence for gas that
can be associated with only the neutral or ionized gas of the Galactic
thick disk.

\section{The Lack of High-Velocity Gas}
\label{sec:highvelocity}

\subsection{High Velocity Clouds}

There is no evidence for discrete high-velocity cloud material along
the \vz\ sight line in \ovi\ or lower-ionization species.  Sembach et
al.  (2003) have discussed the presence of high-velocity \ovi\ toward
extragalactic objects.  They find high-velocity \ovi\ toward
$\sim60\%$ of the sight lines studied (although some were initially
chosen to probe the known \HI\ high-velocity clouds).  In a few cases,
the lack of high-velocity material toward \vz\ is confused by the
presence of high negative-velocity stellar photospheric absorption
features (see, e.g., the \ion{C}{2} 1036.337 \AA\ transition in Figure
\ref{fig:stack}).

For \ovi, our $3\sigma$ upper limit on the column density of
high-velocity ($|\mvlsr|\ga125$ \kms) material is $\log N(\mbox{\ovi})
\la 13.3$, assuming $\sigma \sim 32$ \kms\ ($b\sim45$ \kms).  We note
that $\sim20\%$ of the sight lines studied by Sembach et al.  (2003)
show the presence of very broad, high-positive-velocity wings in \ovi\ 
extending from $|\mvlsr| \sim 100$ to $\sim250$ \kms.  Most of these
wings are found in the northern Galactic sky.  We see no evidence for
low optical depth, high-velocity wings such as these toward \vz.

\begin{figure}
  \epsscale{0.6} 
  \plotone{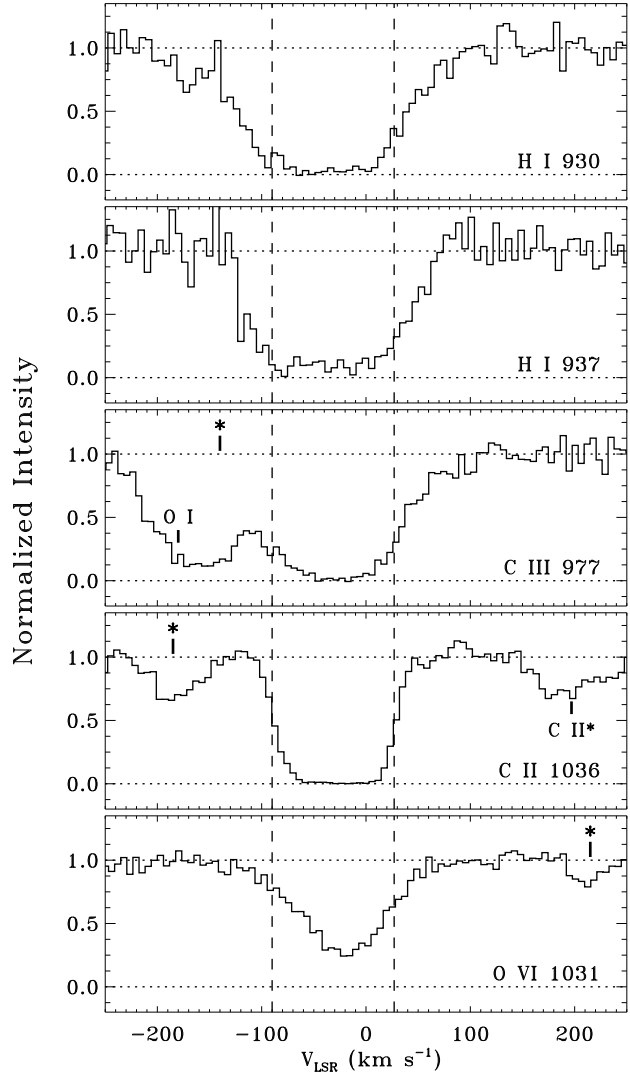}
\caption{Absorption line profiles of \ion{O}{6} $\lambda1031.926$, \ion{C}{2} 
  $\lambda1036.337$, \ion{C}{3} $\lambda 977.020$, \ion{H}{1} $\lambda
  937.803$, and \ion{H}{1} $\lambda 930.748$.  The \ion{H}{1} profiles
  have been normalized by stellar model atmosphere to account for the
  strong, broad stellar photospheric \ion{H}{1} absorption in the
  negative velocity wings of the interstellar profile.  Residual broad
  absorption in these profiles may be in part a result of mismatches
  between the data and the stellar model.  The vertical lines show the
  approximate breadth of the \ion{C}{2} profile at half intensity.  No
  significant high-velocity ($|v_{\rm LSR}| \ga 125$ \kms) absorption
  is seen in any of these profiles, though the negative velocity
  portion of the \HI, \ion{C}{2}, and \ion{C}{3} profiles are strongly
  contaminated by stellar absorption.  The excess flux at the bottom
  of the \ion{H}{1}
  937.803 \AA\ profile is likely residual terrestrial airglow
  emission.
\label{fig:hvc}}
\end{figure}

There are no \HI\ high-velocity cloud complexes toward \vz\ identified
through 21-cm \HI\ emission line observations.  Thus, even though a
large fraction of the extragalactic targets show high-velocity \ovi,
it is unclear if we expect to see such material toward \vz.  To
demonstrate the lack of high-velocity material toward \vz, we show
normalized spectra of two higher-order Lyman series \HI\ lines toward
\vz, together with the profiles of \ion{C}{2} $\lambda 1036.337$,
\ion{C}{3} $\lambda977.020$, and \ovi\ $\lambda 1031.926$ in Figure
\ref{fig:hvc}.  The \HI\ profiles have been normalized by the same
model stellar atmosphere used for the \ion{N}{2} profile (see \S
\ref{sec:analysis}).  Even though this model is a reasonable
descriptor of the broad stellar \HI\ absorption seen in the \fuse\ 
data, we caution that the application of an incorrect model profile
could leave unrealistic broad absorption in the spectrum, particularly
at negative velocities where the absorption from the star is
strongest.  The vertical lines in Figure \ref{fig:hvc} denote the
approximate range of \ion{C}{2} absorption at its half intensity
points.

There is little visible \HI\ beyond the extent of the \ion{C}{2}
profile.  The \HI\ profiles seem to be confined to $-125 \la \mvlsr
\la +65$ \kms\ (with the caveat that the negative velocity extent is
dependent upon the adopted stellar profile).  Outside of this velocity
range, we find an upper limit to the \HI\ column density of any
high-velocity material of $\log N(\mbox{\HI}) \la 14.7$ ($3\sigma$).
Neither do we see evidence for high-velocity absorption in
low-ionization metal species.  Assuming such absorption would have
breadths similar to the values listed in Table \ref{tab:kinematics},
we derive limits of $\log N(\mbox{\feii}) \la 13.4$, $\log
N(\mbox{\fethree}) \la 13.5$, and $\log N(\mbox{\ion{N}{2}}) \la 13.4$
($3\sigma$ limits) for material with $|\mvlsr|\ga125$ \kms.  The
\fethree\ limits only apply to high positive velocities since the
profile is blended with interstellar \feii\ and stellar \fethree\ at
negative velocities.

The intermediate-velocity cloud complex K (Wakker 2002) is found at
lower Galactic latitudes in \HI\ maps at $\mvlsr \approx -70$ \kms.
The map of this complex in Wakker (2002) shows the nearest contour
enclosing $N(\mbox{\HI})>5\times10^{18}$ is $\sim10^\circ$ from \vz.
It is difficult to say if complex K is detected toward \vz.  The very
strongly-saturated \ion{C}{2} 1036.337 \AA\ transition shows some
evidence for weak absorption at velocities consistent with an origin
in complex K.  Small wings at the appropriate velocity may also be
seen in the stronger \ion{O}{1} and \ion{Si}{2} transitions (see
Figure \ref{fig:stack}).  However, the absorption at these velocities
is not sufficiently well delineated to allow us to study it in detail.

\subsection{Globular Cluster Gas}

We note that no absorption associated with the circumstellar
environment of \vz\ or with the globular cluster M 3 is seen in our
data, assuming any gas in the vicinity of M 3 would be found near its
systemic velocity ($\mvlsr = -137.0\pm0.3$; Soderberg et al. 1999).
For \ovi, our $3\sigma$ upper limit on the column density of material
associated with the globular cluster is $\log N(\mbox{\ovi}) \la
13.3$, assuming $\sigma \sim 32$ \kms\ ($b\sim45$ \kms).  Neither do
we see evidence for absorption in lower-ionization species.  Assuming
such absorption would have breadths similar to the values listed in
Table \ref{tab:kinematics}, we derive limits of $\log N(\mbox{\feii})
\la 13.4$ and $\log N(\mbox{\ion{N}{2}}) \la 13.4$ ($3\sigma$).  The
Lyman series of hydrogen cannot be used to search for globular cluster
or circumstellar material because of the strong stellar \HI\ 
absorption.

These determinations can be used to estimate upper limits on the mass
of interstellar material in the globular cluster.  The calculation
follows that for determining the hydrogen column density associated
with \ovi\ in the hot ISM (see \S \ref{sec:osix}) with the additional
step of multiplying that column by a projected area for the cluster.
We assume a metallicity $[{\rm M/H}] \approx -1.57$ (Harris 1996) and
that the upper limits of the above ions are applicable to the whole
cluster.  The distance of \vz\ from the cluster core ($\sim3\farcm9$
or $\sim11$ pc) and the limits on \ovi\ imply an intracluster gas mass
($M_{ICM}$) of $M_{ICM} \la 4.4 \, x(\mbox{\ovi})^{-1}$ \msun.  If we
limit this calculation to temperatures $5.3 \la \log T \la 5.6$ [the
range over which $x(\mbox{\ovi}) \ga 0.05$ for both the equilibrium
calculations of Sutherland \& Dopita (1993) and the non-equilibrium
calculations of Shapiro \& Moore (1976)], we have $M_{ICM} \la 90$
\msun.  Similarly we can use the limits on \ion{N}{2} assuming the
gas is fully ionized (i.e., hydrogen is fully ionized and the
ionization of nitrogen is tied to that of hydrogen via its charge
exchange reaction) to limit the mass of the intracluster medium to be
$M_{ICM} \la 35$ \msun.

Our calculations assume that the limits to the gas column density
toward \vz\ are applicable to the entire projected area within the 11
pc distance to the cluster core.\footnote{We note that M~3 has a
  half-mass radius of $\sim1\farcm1$ or $\sim3.3$ pc and a tidal
  radius of $\sim0\farcm5$ or $\sim1.6$ pc (Harris 1996).}  If the
column of gas is higher toward the center of the cluster or if \vz\ 
lies on the front side of the cluster, then our limits are not
appropriate.  Our limits on the mass of ICM material in M 3 are
similar to other recent limits on the amount of ionized gas in
globular clusters (e.g., Knapp et al. 1996) and the one detection
(Freire et al. 2001).  The density of ionized gas detected in 47~Tuc
by Freire et al. (2002) implies a mass of $\sim10$ \msun\ if the gas
extends as far as 11 pc from the cluster center, the projected
distance of \vz\ from the center of M~3.

\subsection{High-Velocity Dispersion Gas}

Kalberla et al. (1998) have used the LDS \HI\ observations to argue
for the presence of a high velocity-dispersion component of the
Galactic ISM.  They find evidence for diffuse \HI\ with
$\sigma\approx60$ \kms\ and $\log N(\mbox{\HI}) \approx 19.15$.  If
exponentially distributed, Kalberla et al. argue such a distribution
would have a midplane density of $\sim1.2\times10^{-3}$ \percc\ and a
scale height of $\sim4.4$ kpc.


Such a component of the ISM should be readily visible as
strongly-saturated Lyman-series absorption toward distant stars and
AGN.  Savage et al. (2000) have argued on the basis of the measured
equivalent widths of the strong \ion{Mg}{2} doublet near 2800 \AA\ 
that such a phase of the ISM cannot exist unless the abundance of Mg
in this material is far below solar ($\sim 0.05$ solar).

Using the Lyman series of \HI\ as a probe of this gas alleviates the
issue of abundances.  If such material were present toward \vz, the
\HI\ 930.748 \AA\ line seen in Figure \ref{fig:hvc} would have
$\la1\%$ residual intensity over the range $-200 \la \mvlsr \la +200$
\kms.  The \HI\ Lyman series towards \vz\ (conservatively) limits the
column density of such material in this direction to be $\log
N(\mbox{\HI}) \la 16.7$ \column.

While we can say with some confidence that a phase of the ISM which is
smoothly distributed and has a dispersion and column density like
those discussed in Kalberla et al. (1998) does not exist toward \vz,
their detection as a whole cannot be ruled out by the present data
alone.  To identify this phase of the ISM, Kalberla et al. averaged
over $\sim10^\circ \times 10^\circ$ sections of the sky to identify
this gas.  If this high velocity-disperion \HI\ distribution is
contained in clouds which cover a relatively small fraction of the sky
and as an ensemble have a cloud-to-cloud dispersion of order 60 \kms,
we would not necessarily expect to find material associated with this
distribution along a single sight line.  Lockman (2002) has recently
idenfitied a population of discrete thick disk clouds in the inner
Galaxy.  Whether the ensemble of such clouds at the solar circle (if
such a population exists) can match the distribution inferred by
Kalberla et al. (1998) is not known.

An investigation of the validity of the Kalberla et al. (1998) result
is best pursued using a large database of \fuse\ observations of
extragalactic sources (where there is no broad Lyman-series absorption
in the background source itself).  Should no such sight lines through
the entire Galactic halo show evidence for such gas, the reality of
such an \HI\ layer will need to be reevaluated.

\section{Discussion}
\label{sec:discussion}

\subsection{The Phases of the Galactic Thick Disk}

Because we are able to measure the column densities of species
associated with all phases of the ISM toward \vz, we can estimate the
contribution each phase makes to the total hydrogen (neutral plus
ionized) along this sight line.  In this way we can construct a census
of the phases in this direction.  Table \ref{tab:census} gives such a
summary of the phases toward \vz, which contains the hydrogen column
for each phase (as estimated above) and the fraction of the total each
represents along this sight line.  We also summarize the best
estimates of the scale heights of each phase from the literature.

The fraction of the total hydrogen along the \vz\ sight line estimated
to be associated with warm ionized gas is $\sim10\% - 45\%$.  In this
case, the higher end of this range is more robust since it is based
upon a model of the ionization fraction of sulfur in the WIM.  The
lower value assumes that all of the sulfur is in \sthree, which yields
a firm lower limit to the ionized hydrogen column density associated
with the WIM.  This range of fractional hydrogen column density is
similar to the observed values derived by comparing the electron
column densities (derived from pulsar dispersion measures) with \HI\ 
column densities toward distant globular clusters.  Reynolds (1991b)
summarizes five such measurements which give fractions of ionized
hydrogen in the range $19\%-40\%$.

The highly-ionized, transition-temperature gas traced by \ovi\ 
represents a relatively small fraction of the total hydrogen column
density along this path through the Galactic thick disk and corona.
It should be noted that the \ovi\ absorption is likely not tracing
material at X-ray emitting temperatures ($\ga10^6$ K); therefore, the
total amount of hot gas could be higher.  Furthermore, because we have
adopted the highest ionization fraction for estimating the hydrogen
column from \ovi, the stated values should be treated as (uncertain)
lower limits.  


\subsection{The Relationship Between Neutral and Ionized Gas in the
  Galactic Thick Disk}

Our determinations of the depleted gas-phase abundances of iron in the
WNM and WIM are very similar, suggesting the content of iron-bearing
dust is similar in these two phases.  This tends to support the notion
that the gas and dust that make up the WNM and WIM of the Galactic
thick disk have experienced similar dust destruction and formation
histories.  If the processes that feed WNM and WIM material into the
thick disk were significantly different (i.e., if one were more
violent than the other), the dust content, and hence the gas-phase
abundances, of these phases might be significantly different.

We point out, however, that our determination of ${\rm [Fe/S]}_{WIM}$
has a relatively large uncertainty due to the photoionization
modeling.  Using refined models and species that are less susceptible
to changes in the ionizing spectrum (e.g., \ion{Al}{3}; see Howk \&
Savage 1999b) for determining the WIM abundances will provide more
precise results.  It is also worth pointing out that the WNM abundance
determinations may be affected at the $\sim0.1$ dex level by our
neglect of the ionization correction term in Eqn.
(\ref{eqn:abundance}).

The kinematic similarities between the tracers of the WNM (e.g.,
\feii) and the WIM (e.g., \fethree, \sthree) along this sight line
also suggest a relatively close relationship between these phases of
the Galactic thick disk.  The distribution of clouds in velocity space
is likely very similar between the two phases.  We see no significant
difference in the large-scale velocity dispersion of the warm ionized
gas and warm neutral gas toward \vz.  This indicates that these two
phases have similar vertical extents along this high-latitude sight
line, unless processes are at work that affect the vertical
distribution of two phases differently while suppressing large-scale
velocity profile differences.

The kinematic similarities suggest a relationship between these phases
toward \vz\ that is similar to that seen toward HD~93521 (Spitzer \&
Fitzpatrick 1993) and $\rho$~Leo (Howk \& Savage 1999b).  Along these
sight lines, tracers of electron-rich gas (e.g., \ion{C}{2}$^*$,
\ion{Al}{3}, \ion{S}{3}) show a velocity component structure that is
virtually indistinguishable from tracers of the WNM (e.g.,
\ion{Zn}{2}, \ion{Si}{2}, \ion{Fe}{2}).  The apparent similarity of
the \feii\ and \fethree\ profiles as observed by \fuse\ does not rule
out moderate, unresolved component-to-component variations in the
ratio of these species.


The apparent constancy of the \fethree\ to \feii\ ratio with velocity
also implies that there are no strong intermediate- or high-velocity
features with unusual \fethree\ content (i.e., none that are
significantly more highly-ionized than the bulk of the thick disk
material along this sight line).  The resolution of \fuse\ does not
allow us to place strong limits on such components at low velocities,
and the overlap of the strong \feii\ 1121.975 \AA\ transition at
negative velocities relative to \fethree\ also makes the limits more
significant at positive velocities.  

There is one caveat to these kinematic comparisons of the WNM and WIM
tracers.  If the WIM does indeed represent as much as $\sim45\%$ of
the total hydrogen in this direction, much of the column density of
singly-ionized species we have associated with the WNM may arise in
the WIM.  For example, the standard composite WIM model of Sembach et
al. (2000) predicts $x(\mbox{\ion{Fe}{2}}) \approx 0.5$.  If the
fraction of hydrogen toward \vz\ is $\sim50\%$, then $\sim25\%$ of the
\ion{Fe}{2} is expected to arise in the WIM.  In the case of
\ion{Si}{2}, $x(\mbox{\ion{Si}{2}}) \sim 0.9$, implying that $40\%$ of
the column density of this ion could arise in the WIM.  It is not
difficult to understand why the tracers we have naively associated
with the WNM might have very similar kinematics to the intermediate
ions if they are indeed contaminated by this amount of WIM gas.

The observed \ovi\ profile, which traces transition-temperature gas in
the warm-hot ISM, is centered at velocities consistent with the
centroids of the WNM/WIM tracers, but has a much larger velocity
dispersion compared with the tracers of the low and intermediate ions.
The large breadth of the \ovi\ profile may be the result of large
turbulent (or other non-thermal microphysical) motions or large-scale
cloud-to-cloud motions.  The breadth of the \ovi\ profile toward \vz\ 
is consistent with the $N(\mbox{\ovi})$-line width relation identified
by Heckman et al.  (2002).

The agreement between the central velocities of \ovi\ and the lower
ionization species is intriguing, though perhaps coincidental.  The
correspondence of central velocities of all of the tracers of the
thick disk ISM in this direction is naturally explained if the \ovi\ 
arises in interfaces between the warm gas (neutral and ionized) and a
much hotter medium (X-ray emitting material at $T\ga10^6$ K).  Because
of the short cooling time of gas in the range of temperatures at which
\ovi\ is abundant, many classes of theoretical models invoke such an
arrangement (see summary in Spitzer 1996). [Indeed, such an
arrangement calls to mind the clouds envisioned by McKee \& Ostriker
(1977).]  


Examining Figure \ref{fig:hvc} one finds that the total velocity
extent of the strong \ion{C}{2} 1036.337 \AA\ and \ovi\ 1031.926 \AA\ 
profiles is similar.  That is, most of the absorption seen in \ovi\ is
seen at velocities where there is also some lower-ionization material
(as traced by \ion{C}{2}).  The much larger apparent dispersion of the
\ovi\ profile compared with lower-ionization lines of similar peak
optical depths is likely due to components further from the velocity
centroid which exhibit higher \ovi\ to low ion column density ratios.
If one assumes a relatively constant quantity of \ovi\ per interface
(e.g., Shelton \& Cox 1994), then the observed \ovi\ profile traces
the distribution of interfaces as a function of velocity.  The
individual interfaces much overlap in velocity so that they are not
visible as individual components (perhaps being within $\Delta
v\sim2\sigma$ of one another, or $\sim25$ \kms\ for $T\sim3\times
10^5$).

Howk et al. (2002b; see also Howk et al. 2002a) have noted that the
distribution of \ovi\ and \feii\ absorption toward stars in the Large
Magellanic Cloud [$(l,b) \approx (280^\circ, -33^\circ)$] also
requires that the \ovi\ be closely associated with low-ionization
material.  In these directions, the centroids of the Milky Way \ovi\ 
profiles are shifted relative to the \feii\ profiles, although the
total extent of the \ovi\ and \feii\ absorption seem to be well
correlated.  That is, the \ovi\ and \feii\ absorption seemed to trace
many of the same structures, though the changing \ovi /\feii\ ratio
causes the velocity centroids to be different.

If the \ovi\ and low-ionization species trace different portions of
the same structures, the larger dispersion of \ovi\ cannot be simply
explained by a larger temperature in the \ovi -bearing region.
Significant variations must exist in the low-ion to \ovi\ column
density ratio, with the outlying \ovi\ components tracing
higher-ionization gas (i.e., material with a higher \ovi\ to low ion
ratio).  This is required not only so that the velocity dispersion be
lower in the WNM/WIM tracers, but also to be consistent with the
larger \ovi\ scale height compared with the WNM or WIM.  Given the
much larger scale height of the \ovi -bearing gas compared with the
WIM [$h_z(\mbox{\ovi}) \approx2.3 h_z({\rm WIM})$; see Table
\ref{tab:census}], the ratio of \ovi\ to WNM/WIM tracers cannot be
constant with height; a larger ratio is required for material further
from the plane.

Furthermore, if one solely invokes differences in the temperature of
the \ovi -bearing gas, the required temperatures are too high to
maintain large amounts of \ovi.  The largest instrumentally-corrected
dispersion in the profiles of lower-ionization species with peak
optical depths similar to that of \ovi\ (in order to avoid
complications from strong saturation) is that of \sthree: $\sigma_o =
14.7\pm1.7$.  If the observed \ovi\ is associated with the same
components making up the \sthree\ profile, an additional broadening of
$\sigma\sim28$ \kms\ is required to explain the breadth of the \ovi\ 
profile.  This would require $T\sim1.5\times10^6$ K in each component.
At such high temperatures \ovi\ is not expected to be abundant
(Sutherland \& Dopita 1993; Shapiro \& Moore 1976).

The observed arrangement of interstellar absorption toward \vz\ can
perhaps be best explained by a scenario in which \ovi\ and the
lower-ionization species arise in the same structures, with the \ovi\ 
arising in an interface between the warm and hot phases of the ISM.
The material further from the line centroid is found further from the
plane of the Galaxy and is characterized by a higher ionization state
(as one would expect for lower warm material per constant-column \ovi\ 
interface).  In this scenario, the coincidence of the velocity
centroids of \ovi\ and lower-ions requires that most WNM/WIM clouds of
the Galactic thick disk have a corresponding \ovi -bearing interace,
implying that each cloud is embedded within a hot medium.  Further
analysis of the relationship between \ovi\ and the lower-ionization
gas along a larger sample of sight lines is required to determine
whether or not this agreement is simply coincidence.

\subsection{Ionization Corrections to Gas-Phase Abundance
  Determinations}

As discussed in \S \ref{sec:abundances}, the [Fe/H] abundances derived
using the traditional measures (i.e., \feii/\HI) provide a
determination that is within $\sim\pm0.1$ dex of the value derived by
accounting for all of the pertinent ionization stages of iron and
hydrogen [i.e., (\feii\ + \fethree)/(\HI\ + \HII)].  This allowable
range of ionization corrections for determinations of [Fe/H] is
relatively small, implying little ambiguity in studies of gas-phase
abundances in varying Galactic environments.  For iron, which is very
heavily incorporated into dust grains, this ionization correction is
indeed small compared with the typical effects of dust.  Iron is
typically underabundant (depleted) by more than 0.5 dex for WNM
material in the Galactic halo and 1.0 dex for WNM material in the
Galactic disk (Savage \& Sembach 1996). However, it is important to
note that the uncertainties due to the neglect of ionization
corrections are much larger than typical error bars for such
measurements.

The ionization corrections implied here for studying iron in the WNM
agrees well with the limits to the ionization corrections derived by
Sembach \& Savage (1996).  We also note reasonable agreement with the
ionization corrections estimated for the sight line to HD~93521 by
Sembach et al.  (2000), depending on the specific model adopted (see
their Table 7).  However, we caution that the Sembach et al.
discussion of the HD~93521 sight line applies to corrections for the
ions of sulfur; it is not necessarily the case that ionization
corrections derived for iron will apply to the full range of
observable singly-ionized species when compared with \HI.  In
particular, the ionization fraction of \feii\ in the WIM is expected
to be significantly lower than other singly-ionized species such as
\ion{S}{2}, \ion{Si}{2}, and \ion{P}{2} (Sembach et al.  2000).  Thus,
the ionization corrections for these species may be larger than that
for iron.

More extensive studies of the ionization corrections along
high-latitude sight lines would clearly be useful in determining the
applicability of our values to broader studies of the Galactic thick
disk.

\section{Summary}
\label{sec:summary}

We have presented {\em Far Ultraviolet Spectroscopic Explorer}
observations of the distant post-AGB star von~Zeipel 1128, which
resides in the globular cluster Messier 3.  We have derived column
densities and kinematic information for the prominent phases of the
ISM along this sight line using the absorption lines present in the
\fuse\ data.  The major conclusions of this work are as follows.

\begin{enumerate}
  
\item Thick disk material toward \vz\ is centered near $\mvlsr \sim
  -26$ \kms, implying infall of this material onto the Galactic plane
  in this general direction.  Species tracing the prominent warm
  neutral, warm ionized, and transition temperature ionized phases are
  seen at very similar velocities along this sight line.
  
\item As much as $\sim45\%$ of the hydrogen toward \vz\ may be
  associated with the WIM.  The tracers of WNM and WIM gas toward \vz\ 
  have very similar kinematic profiles, showing both the same central
  velocity and very similar breadths.
  
\item After accounting for all of the ionization stages of iron and
  hydrogen, we find that the ionization corrections to the standard
  method of determining the gas-phase abundance of [Fe/H] are likely
  of order 0.1 dex.
  
\item The gas-phase abundance of [Fe/S] in the WIM is estimated by
  comparing the ratio \fethree/\sthree\ and applying an ionization
  correction (using the models of Sembach et al. 2000).  The gas-phase
  abundance of iron in the WIM is indistinguishable from that of the
  WNM toward \vz.  The WIM gas-phase abundance of iron, $[{\rm
    Fe/S}]_{WIM} \approx -0.74\pm0.3$, suggests significant
  incorporation of iron into dust in this phase of the Galactic thick
  disk.
  
\item The similarities of the kinematics and gas-phase abundances of
  the WNM and the WIM suggest the two phases of the Galactic thick
  disk are closely related along this sight line.  The WNM and WIM in
  this direction likely have similar vertical extents and similar
  Fe-bearing dust content.
  
\item The broad ($\sigma = 32$ \kms) \ovi\ absorption toward \vz\ is
  centered at the same velocities as the WNM and WIM tracers along
  this sight line.  The low- and high-ionization gas along this sight
  line seem to be closely related given the similarity in their
  central velocities and total velocity extent.  It seems likely that
  the \ovi\ arises in interfaces between hot and warm gas, with the
  shape of the \ovi\ profile tracing the distribution of interfaces in
  velocity.
  
\item We see no high-velocity material that could be associated with
  the circumstellar environment of \vz\ or with the globular cluster
  Messier 3 in which \vz\ resides.  No high-velocity clouds are seen
  along this sight line in the low- or high-ionization states observed
  by \fuse.  No high-velocity dispersion gas is seen in the \HI\ 
  Lyman-series absorption.

\end{enumerate}

\acknowledgements

We thank P. Chayer for calculating a model stellar atmosphere for
comparison with the \fuse\ observations of \vz.  We make use of data
from the Wisconsin H-Alpha Mapper, which is funded by the National
Science Foundation.  This work is based on data obtained for the
Guaranteed Time Team by the NASA-CNES-CSA FUSE mission operated by the
Johns Hopkins University. Financial support to U. S. participants has
been provided by NASA contract NAS5-32985.  JCH and KRS recognize
support from NASA Long Term Space Astrophysics grant NAG5-3485 through
The Johns Hopkins University.




\begin{deluxetable}{lrc}
\tablenum{1}
\tablecolumns{3}
\tablewidth{0pt}
\tablecaption{Properties of von Zeipel 1128
\label{tab:star}}
\tablehead{
\colhead{Property} & 
\colhead{Value} &
\colhead{Reference\tablenotemark{a}}
}
\startdata
Cluster         & M~3 (NGC~5272)        & \nodata \\
RA (J2000)      & 13 42 16.84           & \nodata\\
Dec. (J2000)    & $+28$ 26 00.8         & \nodata \\
$(l,b)$         & $(42\fdg5,\ +78\fdg7)$ & \nodata \\
Distance [kpc]  & 10.2                  & 1 \\
$z$ [kpc]       & 10.0                  & 1 \\
Sp. Type        & O8p                   & 2 \\
V [mag.]        & 14.93                 & 3 \\
B-V [mag.]      & $-0.27$               & 3 \\
$T_{eff}$ [K]   & $35,000\pm1,000$      & 4 \\
$\log g$        & $4.0\pm0.25$          & 4 \\
$\log L/L_\odot$ & $3.21\pm0.12$        & 4 \\
$v_{\rm LSR, cluster}$ [\kms]
                & $-137.0\pm0.3$        & 5 \\
$v_{\rm LSR, vZ1128}$ [\kms]
                & $-140\pm8$            & 6 \\ 
\enddata
\tablenotetext{a}{References: (1) Djorgovski 1993; (2) Garrison \& 
Albert 1986; (3) Johnson \& Sandage 1956; (4) Dixon et al. 1994; 
(5) Soderberg et al. 1999; (6) this work.}
\end{deluxetable}


\pagebreak


\begin{deluxetable}{llcr}
\tablenum{2}
\tablecolumns{4}
\tablewidth{0pt}
\tablecaption{Log of \fuse\ Observations
\label{tab:log}}
\tablehead{
\colhead{Visit} &
\colhead{Date} & 
\colhead{No. of} &
\colhead{Exp. Time} 
\\
\colhead{ID} &
\colhead{} &
\colhead{Exp.} &
\colhead{[sec]} 
}
\startdata
P1014101 & 6/18/2000 & 7 & 9119 \\
P1014102 & 6/19/2000 & 9 & 7684 \\
P1014103 & 6/22/2000 & 9 & 14723 \\
\cline{0-3}
\multicolumn{2}{r}{Total:} & 25 & 31526 \\
\enddata
\end{deluxetable}

\pagebreak


\begin{deluxetable}{llcc}
\tablenum{3}
\tablecolumns{4}
\tablewidth{0pt}
\tablecaption{Stellar Radial Velocity Measurements\tablenotemark{a}
\label{tab:stellarvelocities}}
\tablehead{
\colhead{Species} &
\colhead{$\lambda$} & \colhead{Ref.\tablenotemark{b}}  &
\colhead{$v_{\rm LSR}$ [km s$^{-1}$]}}
\startdata
\ion{P}{4} & 1030.515 & 1 & $-140.7\pm2.3$ \\
\ion{P}{4} & 1033.112 & 1 & $-135\pm7$ \\
\ion{P}{4} & 1035.516 & 1 & $-136\pm3$ \\
\ion{S}{4} & 1062.678 & 2 & $-142.1\pm1.1$ \\
\ion{S}{4} & 1072.996 & 2 & $-138.3\pm1.1$ \\
\ion{S}{4} & 1073.528 & 2 & $-139.1\pm1.0$ \\
\ion{P}{5} & 1117.977 & 3 & $-141.6\pm1.5$ \\
\ion{P}{5} & 1128.008 & 3 & $-141.6\pm1.3$ \\
\ion{C}{3} & 1174.933 & 4 & $-139.2\pm1.4$ \\
\ion{C}{3} & 1176.370 & 4 & $-141.1\pm1.3$ \\
\enddata
\tablenotetext{a}{We have chosen only cleanly-separated transitions in
  the LiF1A and LiF1B spectra for these measurements.}
\tablenotetext{b}{Wavelength references: (1) Zetterberg \& Magnusson
        1977; (2) Irwin \& Livingston 1976; (3) Magnusson \&
        Zetterberg 1974; (4) Moore 1970.}

\end{deluxetable}

\pagebreak


\begin{deluxetable}{lrccccccc}
\tabletypesize{\small}
\tablenum{4}
\tablecolumns{9}
\tablewidth{0pt}
\tablecaption{Selected Interstellar Absorption Lines Towards 
        vZ~1128\tablenotemark{a}
\label{tab:measurements}}
\tablehead{
\colhead{Species} & \colhead{$\lambda$} &
\colhead{$\log \lambda f$} &
\multicolumn{2}{c}{$W_\lambda$ [m\AA]} & &
\multicolumn{2}{c}{$\log N_a(v)$} &
\colhead{$v_-,v_+$} \\
\cline{4-5} \cline{7-8}
\colhead{} & \colhead{[\AA]} &
\colhead{} & 
\colhead{[Det 1]} & \colhead{[Det 2]} & &  
\colhead{[Det 1]} & \colhead{[Det 2]} & 
\colhead{[km s$^{-1}$]}
}
\startdata
\ion{C}{2} & 1036.337 & 2.106  &
        $399\pm5$ & $400\pm7$ & & $>14.55$ & $>14.55$ & $-120,+45$ \\
%
\ion{C}{3} &  977.020 & 2.872 &
        $457\pm9$ & $452\pm7$ & & $>13.85$ & $>13.85$ & $-100,+120$ \\
\ion{N}{1} & 964.626 &  0.959 &
   $110\pm6$ & $115\pm5$ & & $15.28\pm0.03$ & $15.35\pm0.03$ & $-65,+45$ \\
\ion{N}{1} & 1134.165 & 1.238 &
        $154\pm4$\tablenotemark{b} & $172\pm3$\tablenotemark{b} &  &
        $15.22\pm0.03$ & $15.33\pm0.03$ & $-55,+50$ \\
\ion{N}{1} & 1134.415 & 1.528 &
  $207\pm5$ & $204\pm4$ & & $15.06\pm0.03$ & $15.15\pm0.04$ & $-50,+50$ \\
\ion{N}{1} & 1134.980 & 1.693 &
  $220\pm6$ & $231\pm4$ & & $14.94\pm0.04$ & $15.08\pm0.18$ & $-95,+45$ \\
\ion{N}{2}\tablenotemark{c} & 1083.994 & 2.100 &
  $393\pm11$ & $393\pm13$ & & $>14.52$ & $>14.52$ & $-110,+55$ \\
\ion{O}{1} & 924.950 & 0.155 & 
    $134\pm9$ & $145\pm7$ & & $16.25\pm0.04$ & $16.37\pm0.03$ & $-65,+40$ \\
\ion{O}{1} & 929.517 & 0.329 & 
    $150\pm12$ & $167\pm7$ & & $16.16\pm0.04$ & $16.25\pm0.03$ & $-95,+50$\\
\ion{O}{1} & 950.885 & 0.176 & 
    $146\pm8$ & $157\pm6$ & & $16.23\pm0.03$ & $16.34\pm0.02$ & $-65,+30$ \\
\ion{O}{1} & 1039.230 & 0.980 &
        $249\pm4$ & $245\pm5$ & & $>15.89$ & $>15.84$ & $-85,+40$ \\
\ion{O}{6} & 1031.926 & 2.137 &
    $260\pm7$ & $263\pm8$ & & $14.49\pm0.03$ & $14.51\pm0.03$ & $-160,+100$ \\
\ion{O}{6} & 1037.617 & 1.836 & 
     $150\pm6$ & $150\pm7$ & & $14.48\pm0.03$ & $14.47\pm0.03$ & $-95,+50$ \\
\ion{Si}{2} & 1020.699 & 1.460 &
   $153\pm4$ & $158\pm6$ && $15.20\pm0.02$ & $15.22\pm0.03$ & $-80,+30$ \\
\ion{P}{2} & 1152.818 & 2.435 &
        $38\pm6$\tablenotemark{d} & $60\pm4$\tablenotemark{d}  
        & & $13.20\pm0.07$\tablenotemark{d} & $13.41\pm0.04$\tablenotemark{d} 
        & $-75,+30$ \\
\ion{P}{3} & 998.000 & 2.047 &
        $<34$ & $<25$ & & $<13.53$ & $<13.41$
        & \nodata \\
\ion{P}{5} & 1117.977 & 2.723 &
        $<25$ & $<25$ & & $<12.68$ & $<12.68$
        & \nodata \\
\ion{S}{3} & 1012.495 & 1.556 &
        $92\pm4$ & $91\pm6$ & & $14.47\pm0.02$ & $14.47\pm0.03$ & $-90,+20$ \\
\ion{S}{4} & 1062.664 & 1.799 &
        $<28$ & $<34$ & & $<13.67$ & $<13.75$
        & \nodata \\
\ion{S}{6} & 944.523 & 2.311 &
        $<43$ & $<45$ & & $<13.40$ & $<13.40$
        & \nodata \\
\ion{Ar}{1} & 1048.220 & 2.408 & 
        $90\pm4$ & $97\pm4$ & & $13.70\pm0.03$ & $13.71\pm0.02$ & $-70,+25$ \\
\ion{Ar}{1} & 1066.660 & 1.851 & 
        $51\pm3$ & $44\pm5$ & & $13.94\pm0.04$ & $13.88\pm0.05$ & $-55,+40$ \\
\ion{Fe}{2} & 1055.262 & 0.898 & 
        $54\pm4$ & $49\pm5$ & & $14.92\pm0.03$ & $14.88\pm0.04$ & $-90,+15$ \\
\ion{Fe}{2} & 1063.176 & 1.765 & 
        $165\pm5$ & $160\pm6$ & & $14.73\pm0.03$ & $14.68\pm0.04$ & $-90,+35$ \\
\ion{Fe}{2} & 1096.877 & 1.545 & 
        \tablenotemark{e} & $132\pm4$ &  &\tablenotemark{e} & $14.76\pm0.02$ 
        & $-75,+25$ \\
\ion{Fe}{2} & 1125.448 & 1.255 & 
        $93\pm4$ & $93\pm4$ & & $14.83\pm0.02$ & $14.83\pm0.02$ & $-75,+35$ \\
\ion{Fe}{2} & 1127.098 & 0.483 & 
        $14\pm3$ & $13\pm3$ & & $14.69\pm0.09$ & $14.65\pm0.09$ & $-55,+5$ \\
\ion{Fe}{2} & 1142.366 & 0.681 & 
        $24\pm4$ & $27\pm3$ & & $14.75\pm0.08$ & $14.79\pm0.05$ & $-50,+20$ \\
\ion{Fe}{2} & 1143.226 & 1.306 & 
        $96\pm5$ & $90\pm4$ & & $14.78\pm0.03$ & $14.77\pm0.02$ & $-80,+25$ \\
\ion{Fe}{2} & 1144.938 & 1.978 & 
        $231\pm6$ & $235\pm4$ & & $14.68\pm0.06$ & $14.74\pm0.04$ & $-95,+30$ \\
\ion{Fe}{3} & 1122.524 & 1.786 & 
        $115\pm4$ & $115\pm4$ & & $14.39\pm0.02$ & $14.42\pm0.02$ & $-90,+15$ \\
%
%
\enddata
\tablenotetext{a}{Measurements are given for \fuse\ detectors 1 and 2.
  Values refer to measurements with LiF1/LiF2 for $\lambda>1000$ \AA\
  and SiC1/SiC2 for $\lambda<1000$. All upper limits are $3\sigma$
  estimates and were derived assuming the transition has a breadth
  equal to the observed breadths for transitions of similar species
  (see \S 4).  Rest wavelengths are from D. Morton 1999, private
  communication.}
\tablenotetext{b}{Line contaminated by detector fixed-pattern noise.}
\tablenotetext{c}{The \ion{N}{2} result is based on measurements of
  SiC1A and SiC2B data.  Due to the overlap of the interstellar line
  by a broad stellar feature, we have used a model atmosphere kindly
  calculated by P. Chayer 2001, private communication, to normalize
  the spectrum before measuring this line (see text).  The model is an
  excellent fit to the stellar lines in this region.}
\tablenotetext{d}{The measurements of \ion{P}{2} in the two detectors
  are not in agreement, likely due to detector effects.  We adopt the
  LiF2 measurement, but do not place too much confidence in it (see
  text).}
\tablenotetext{e}{Line lies near the edge of the detector and
  therefore no accurate measurement is possible.}
\end{deluxetable}

\pagebreak


\begin{deluxetable}{lccc}
\tabletypesize{\small}
\tablenum{5}
\tablecolumns{3}
\tablewidth{0pt}
\tablecaption{Adopted Interstellar Column Densities 
\label{tab:columns}}
\tablehead{
\colhead{Species} & 
\colhead{$\log N$\tablenotemark{a}} &
\colhead{$b$ [\kms]\tablenotemark{b}} &
\colhead{Method\tablenotemark{c}}
}
\startdata
\ion{H}{1}  & $19.97\pm0.03$    & \nodata       & \tablenotemark{d} \\
H$_2$       & $<14.35$          & \nodata       & AOD\tablenotemark{e} \\
\ion{C}{2}  & $>14.55$          & \nodata       & AOD \\
\ion{C}{3}  & $>13.85$          & \nodata       & AOD \\
\ion{N}{1}  & $>15.35$          & \nodata       & AOD \\
\ion{N}{2}  & $>14.52$          & \nodata       & AOD \\
\ion{O}{1}  & $>16.35$          & \nodata       & AOD \\
\ion{O}{6}  & $14.49\pm0.03$    & \nodata       & AOD \\
\ion{Si}{2} & $>15.20$          & \nodata       & AOD \\
\ion{P}{2}  & $13.46\pm0.10$    & $>9.9$\tablenotemark{f}       
                                                & AOD,COG\tablenotemark{f} \\
\ion{P}{3}  & $<13.4$           & \nodata       & AOD \\
\ion{P}{5}  & $<12.7$           & \nodata       & AOD \\
\ion{S}{3}  & $14.47\pm0.03$    & \nodata       & AOD \\
\ion{S}{4}  & $<13.7$           & \nodata       & AOD \\
\ion{S}{6}  & $<13.4$           & \nodata       & AOD \\
\ion{Ar}{1} & $14.02\pm0.05$    & $9.9\pm0.8$   & COG \\
\ion{Fe}{2} & $14.80\pm0.05$    & $22.9\pm1.0$  & COG   \\
\ion{Fe}{3} & $14.42\pm0.05$    & $22.9\pm2.0$\tablenotemark{g} 
                                                & AOD,COG \\
%
\enddata
\tablenotetext{a}{Adopted column density.  All upper limits are
  $3\sigma$ estimates.}
\tablenotetext{b}{Doppler parameter derived via curve-of-growth
  analysis.
%
%
  This quantity is not given for species for which the column
  densities were derived via the apparent optical depth.}
\tablenotetext{c}{AOD: Apparent optical depth method.  Column
  densities are derived from direct integrations of the apparent
  column density profiles. COG: Curve-of-growth fit to measured
  equivalent widths of the lines for each species.}
\tablenotetext{d}{\ion{H}{1} column density derived from STIS echelle
  observations of \lya\ from Howk et al. 2003.}
\tablenotetext{d}{The limits on H$_2$ are derived in \S 3.}
\tablenotetext{f}{The adopted \ion{P}{2} column density is an average
  of the values derived using apparent column density and curve of
  growth analyses of the LiF2A data.  The latter assumes a $b$-value
  equivalent to that derived for \ion{Ar}{1}.  Given the potential
  systematic uncertainties, the quoted uncertainties represent the
  linear combination of the uncertainties in each method.}
\tablenotetext{g}{The \ion{Fe}{3} column density has been derived
  using a curve of growth with the $b$-value (with twice the
  uncertainties) derived for \ion{Fe}{2}.  The AOD method gives
  consistent results.}
\end{deluxetable}

\pagebreak


\begin{deluxetable}{lcccc}
\tablenum{6}
\tablecolumns{4}
\tablewidth{0pt}
\tablecaption{Kinematic Properties of Select Species
\label{tab:kinematics}}
\tablehead{
\colhead{Species} & 
\colhead{$\langle v_{\rm LSR} \rangle$\tablenotemark{a}} &
\colhead{$\langle \sigma_{obs} \rangle$\tablenotemark{b}} &
\colhead{$\langle \sigma_{o} \rangle$\tablenotemark{c}} &
\colhead{$\langle \tau_{max} \rangle$\tablenotemark{d}} \\
\colhead{} & 
\colhead{[km s$^{-1}$]} & \colhead{[km s$^{-1}$]} & 
\colhead{[km s$^{-1}$]} & \colhead{}
} 
\startdata 
%
\ion{O}{6}  & $-26.3\pm1.3$ & $32.9\pm1.1$ & $31.8\pm1.1$ & 1.01 \\ 
\ion{Si}{2} & $-20.7\pm0.5$ & $18.2\pm0.4$ & $16.1\pm0.4$ & 1.55 \\ 
\ion{P}{2}  & $-22\pm4$     & $14\pm3$     & $11\pm3$     & 0.58 \\ 
\ion{S}{3}  & $-26.8\pm2.1$ & $17.0\pm1.7$ & $14.7\pm1.7$ & 0.79 \\ 
\ion{Ar}{1} & $-9.7\pm0.6$  & $13.7\pm0.5$ & $10.7\pm0.5$ & 0.94 \\ 
\ion{Fe}{2} & $-23.4\pm1.1$ & $14.2\pm0.9$ & $11.4\pm0.9$ & 0.80 \\ 
\ion{Fe}{3} & $-30.8\pm0.8$ & $15.3\pm0.6$ & $12.7\pm0.6$ & 0.95 \\ 
\enddata
\tablenotetext{a}{Average of the central velocities determined by
  fitting Gaussian models to the $N_a(v)$ profiles.  The uncertainties
  are statistical only (they do not include the relative uncertainties
  associated with the {\em FUSE} wavelength solution or the absolute
  correction to the LSR frame discussed in the text).}
\tablenotetext{b}{Average of the fitted Gaussian dispersions.}
\tablenotetext{c}{Gaussian dispersions corrected for the contribution
  of the instrumental line spread function, assumed to be
  $\sigma_{LSF} \approx 8.5$ km s$^{-1}$.}  \tablenotetext{d}{Average
  peak optical depth of the transitions used in the Gaussian fits.}
\end{deluxetable}


\begin{deluxetable}{lcc}
\tablenum{7}
\tablecolumns{3}
\tablewidth{0pt}
\tablecaption{WNM Gas-phase Abundances
\label{tab:wnmabundances}}
\tablehead{
\colhead{Species} & 
\colhead{$\log (X/{\rm H})_\odot+12$\tablenotemark{a}} &
\colhead{$[X/{\rm H}]$\tablenotemark{b}}
}
\startdata
\ion{C}{2}  & 8.52 & $>-1.94$ \\
\ion{N}{1}  & 7.93 & $>-0.55$ \\
\ion{O}{1}  & 8.74 & $>-0.36$ \\
\ion{Si}{2} & 7.55 & $>-0.32$ \\
\ion{P}{2}  & 5.56 & $-0.07\pm0.10$ \\
\ion{Ar}{1} & 6.48 & $-0.43\pm0.06$\tablenotemark{c} \\
\ion{Fe}{2} & 7.50 & $-0.64\pm0.06$ \\
\enddata
\tablenotetext{a}{Solar system abundances adopted from Grevesse \&
  Sauval 1998 with the following exceptions. Argon: We adopt the Sofia
  \& Jenkins 1998 value, which is the average of solar and local
  B-star photospheric abundances.  Oxygen and nitrogen: we adopt the
  new photospheric determinations of Holweger 2001. }
\tablenotetext{b}{Normalized gas-phase abundance, relative to
  hydrogen: $[X/{\rm H}] \equiv \log [N(X)/N(\mbox{\ion{H}{1}})]-\log
  (X/H)_\odot$.  This treatment neglects ionization effects as well as
  the effects of the large beam used for determining
  $N(\mbox{\ion{H}{1}})$.}
\tablenotetext{c}{The apparent deficiency of \ion{Ar}{1} is likely
  caused by the over-ionization of \ion{Ar}{1} relative to \ion{H}{1}.
  See Sofia \& Jenkins 1998 for a discussion of this effect.}
\end{deluxetable}


\begin{deluxetable}{lccccl}
\tablenum{8}
\tablecolumns{6}
\tablewidth{0pt}
\tablecaption{Census of Phases Toward von Zeipel 1128
\label{tab:census}}
\tablehead{
\colhead{Phase} &
\colhead{Scale Height} &
\colhead{Ref.\tablenotemark{a}} &
\colhead{$\log N({\rm H})$\tablenotemark{b}} &
\colhead{Fraction\tablenotemark{c}} &
\colhead{Tracer\tablenotemark{d}}}
\startdata
Warm Neutral & 0.4 kpc\tablenotemark{e} 
                       & 1,2 & 19.97       & $53\% - 86\%$ & \ion{H}{1} \\
Warm Ionized & 1.0 kpc & 3,4 & $19.1-19.9$ & $12\% - 45\%$ & \ion{S}{3} \\
Warm-Hot Ionized  & 2.3 kpc & 5   & $\ga18.4$   & $\ga1\% - 2\%$ & \ion{O}{6} \\
\enddata
\tablenotetext{a}{Reference for scale height determinations:
        (1) Savage \& Massa 1987; (2) Diplas \& Savage 1994; (3)
        Haffner et al. 1999; (4) Savage, Edgar, \& Diplas 1990; (5)
        Savage et al. 2002.}
\tablenotetext{b}{Estimated total hydrogen column density associated 
        with each phase of the ISM (see text).}
\tablenotetext{c}{Fraction of total hydrogen column density associated 
        with each phase of the ISM.  A range of values is given to
        account for the various ionization corrections assumed.}
\tablenotetext{d}{Tracer used to derive the hydrogen columns associated
        with each phase of the ISM.}
\tablenotetext{e}{The \ion{H}{1} scale height given here corresponds to
        the extended thick disk distribution rather than the thin disk
        distribution (see, e.g., Diplas \& Savage 1994; Dickey \&
        Lockman 1990).}
\end{deluxetable}




\end{document}